\documentclass[aps,twocolumn,pre,floatfix]{revtex4-2}

\usepackage{xcolor,hyperref}
\hypersetup{
   colorlinks,
   linkcolor={blue!50!black},
   citecolor={blue!50!black},
   urlcolor={blue!80!black}
}

\usepackage{epsfig, amsmath}
\usepackage{bm}
\usepackage{soul,ulem}
\usepackage{hyperref}
\usepackage{footmisc}
\usepackage{wrapfig}
\usepackage{amssymb}
\DeclareMathAlphabet{\mathitbf}{OML}{cmm}{b}{it}
\newcommand{\nav}{u_{\mbox{\scriptsize na}}^2}

\newcommand{\dv}{\mathitbf d}

\newcommand{\uv}{\mathitbf u}
\newcommand{\xv}{\mathitbf x}
\newcommand{\yv}{\mathitbf y}

\newcommand{\nv}{\mathitbf n}

\newcommand{\calBold}[1]{\mbox{\boldmath${\cal #1}$}}
\newcommand{\sFrac}[2]{{\textstyle\frac{#1}{#2}}}
\newcommand{\dbar}{{\,\mathchar'26\mkern-12mu d}}

\definecolor{darkGreen}{RGB}{0,100,0}

\begin{document}

\title{Effects of coordination and stiffness scale-separation in disordered elastic networks}
\author{Edan Lerner}
\email{e.lerner@uva.nl}
\affiliation{Institute for Theoretical Physics, University of Amsterdam, Science Park 904, 1098 XH Amsterdam, the Netherlands}

\begin{abstract}
Many fibrous materials are modeled as elastic networks featuring a substantial separation between the stiffness scales that characterize different microscopic deformation modes of the network's constituents. This scale separation has been shown to give rise to emergent complexity in these systems' linear and nonlinear mechanical response. Here we study numerically a simple model featuring said stiffness scale-separation in two-dimensions and show that its mechanical response is governed by the competition between the characteristic stiffness of collective nonphononic soft modes of the stiff subsystem, and the characteristic stiffness of the soft interactions. We present and rationalize the behavior of the shear modulus of our complex networks across the unjamming transition at which the stiff subsystem alone loses its macroscopic mechanical rigidity. We further establish a relation in the soft-interaction-dominated regime between the shear modulus, the characteristic frequency of nonphononic vibrational modes, and the mesoscopic correlation length that marks the crossover from a disorder-dominated response to local mechanical perturbations in the near-field, to a linear, continuum-like response in the far field. The effects of spatial dimension on the observed scaling behavior are discussed, in addition to the interplay between stiffness scales in strain-stiffened networks, which is relevant to understanding the nonlinear mechanics of non-Brownian fibrous biomatter.


\end{abstract}

\maketitle

\section{Introduction}

Disordered networks of unit masses connected by Hookean springs constitute a popular minimal model for studying the elasticity of disordered solids~\cite{shlomo,matthieu_thesis,mw_maha_prl_2008,Ellenbroek_2009,mw_EMT_epl,chase_nature_physics_2011,phonon_gap_2012,Xiaoming_emt_2013,Xiaoming_kagome_2013,sss_epje_2018}. In this simple model the key parameter that controls macro- and microscopic elasticity is the mean node coordination $z$. Several scaling laws have been established for characteristic length-scales~\cite{breakdown,brian_prl_2017,sss_epje_2018}, frequency-scales~\cite{matthieu_thesis,mw_EMT_epl,phonon_gap_2012}, elastic moduli~\cite{matthieu_thesis,Ellenbroek_2009} and their fluctuations~\cite{chi_paper_2023,giannini2024scaling} in terms of the difference $\delta z\!\equiv\!z\!-\!z_{\rm c}$, where $z_{\rm c}\!\equiv\!2\dbar$ is the Maxwell threshold, and $\dbar$ stands for the dimension of space. 

Inspired by observations from rheological experiments on biomaterials~\cite{PhysRevLett.75.4425,PhysRevE.82.051905}, the simple disordered spring-network model in two dimensions (2D) was supplemented by \emph{bending} (angular) interactions that penalize changes in the angles formed between edges connected to the same node~\cite{fred_prl_2003,chase_nature_physics_2011,RevModPhys.86.995,mackintosh_prl_2019}. Particularly relevant to investigating fibrous biomaterials is the case in which the stiffness associated with these angular interactions is far smaller than the stiffness associated with the stretching or compression of the radial, Hookean springs. In this limit, the phenomenon of strain-stiffening of athermal, floppy ($z\!<\!z_{\rm c}$) elastic networks stabilized by soft angular interactions was thoroughly investigated in recent years~\cite{gustavo_pre_2014,feng_soft_matter_2016,robbie_nature_physics_2016,mackintosh_pre_2016,robbie_pre_2018,merkel_pnas_2019,mackintosh_prl_2019,fred_prl_2023,strain_stiffening_2023}. Despite the aforementioned efforts and progress, the combined effects of changing \emph{both} coordination $z$ \emph{and} the ratio $\kappa$ of bending-to-stretching stiffnesses of disordered elastic networks, have not been fully resolved. 

\begin{figure}[ht!]
\includegraphics[width = 0.5\textwidth]{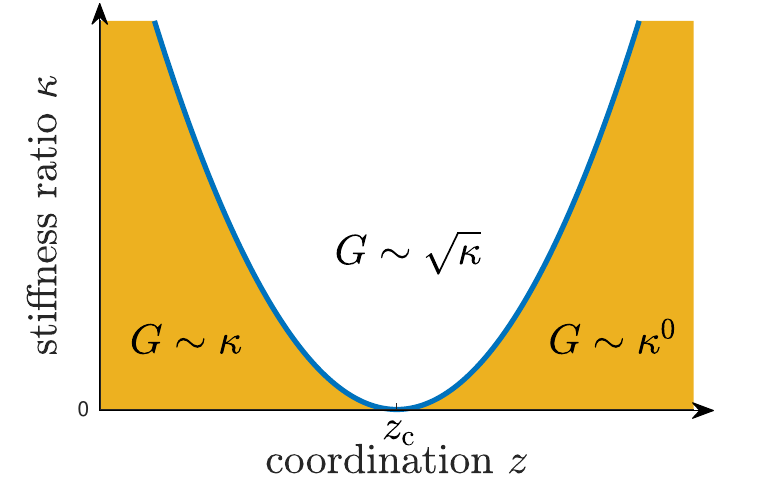}
  \caption{\footnotesize Scaling regimes of the shear modulus $G$ in disordered networks characterized by their coordination $z$ and the bending-to-stretching stiffness ratio $\kappa$. The crossover between the small- and large-$\kappa$ scaling occurs when $\kappa\!\sim\!(z\!-\!z_{\rm c})^2$ (with logarithmic corrections in two-dimensions), see text for discussion. We note that the above illustration described the mechanics near $z\!=\!z_{\rm c}$ and $\kappa\!\ll\!1$; different scaling laws are expected away from these regimes, cf.~Refs.~\cite{chase_nature_physics_2011,Xiaoming_emt_2013,Xiaoming_kagome_2013}.}
  \label{fig:fig1}
\end{figure}

In this work we fill this gap and study the mechanics of disordered spring networks endowed with soft, angular interactions, under variations of both the coordination $z$ of the underlying stiff network, and of the ratio $\kappa\!\ll\!1$ of bending-to-stretching stiffness. We trace out the crossover in our networks' shear modulus between a regime dominated by the stiff subnetwork's nonphononic soft (for $z\!>\!z_{\rm c}$) or zero (for $z\!<\!z_{\rm c}$) modes, and a regime dominated by the soft angular interactions, as illustrated in Fig.~\ref{fig:fig1}. Interestingly, we find that, in both regimes, the shear modulus $G$ scales with the characteristic \emph{frequency} of soft, nonphononic modes, independent of whether the latter are dominated by the stiff subnetwork, or by the soft angular interactions, in states for which the stiff subnetwork alone has a finite shear modulus. In addition, we show that the same scaling argument made in Ref.~\cite{robbie_epje_2019} for the mesoscopic correlation length $\xi$ --- that marks the crossover between a non-affine, near-field disorder-dominated response to local perturbations, to an affine, far-field linear-continuum-elastic-like response ---, holds in the unexplored, angular-interaction-dominated regime. We further show that scaling laws for shear moduli in 2D feature logarithmic corrections that are absent in 3D. Finally, we investigate the effects of stiffness scale separation in strain-stiffened networks in 2D.

This work is structured as follows; in Sect.~\ref{sec:model_and_methods} we present the models and methods employed, and the precise definitions of the key observables considered in this work. In Sect.~\ref{sec:results} we present our numerical results for the elastic properties of isotropic disordered networks and rationalize them using scaling arguments, building on existing arguments and results. In Sect.~\ref{sec:strain_stiffened} we show results and discuss the scaling behavior seen for strain-stiffened elastic networks. We summarize this work and discuss future research questions in Sect.~\ref{sec:summary}.


\section{Numerical model, methods and observables}
\label{sec:model_and_methods}

We employ a well-studied model of athermal biopolymer fibrous materials~\cite{fred_prl_2003,chase_nature_physics_2011,RevModPhys.86.995,mackintosh_prl_2019}. Given a disordered, two-dimensional (2D) graph of nodes and edges with some connectivity $z$ (see below how graphs were obtained in this study), we assign a Hookean spring with stiffness $k_r$ (set to unity in our calculations) at its rest length $\ell$ on each edge. Every two edges emanating from the same node --- with no other edge between them --- form angles $\theta$; we assign an angular spring at its rest-angle $\theta^{(0)}$ to each such angle, with stiffness $k_\theta$ that has units of energy. The potential energy of this model therefore reads
\begin{equation}\label{eq:potential_energy}
    U = \sFrac{1}{2}k_r\!\!\!\!\sum_{\mbox{\scriptsize edges }ij}\!\!\!\!(r_{ij}-\ell_{ij})^2 + \sFrac{1}{2}k_\theta\!\!\!\!\!\sum_{\mbox{\scriptsize angles }ijk}\!\!\!\!\!(\theta_{ijk}-\theta^{(0)}_{ijk})^2\,.
\end{equation}
In what follows we express all energies in terms of $k_r\bar{\ell}^2$ and elastic moduli in terms of $k_r\bar{\ell}^{2-\dbar}$ where $\bar{\ell}\!\equiv\!(V/N)^{1/\dbar}$ with $V$ denoting the system's volume, $N$ denoting the number of nodes, and $\dbar$ stands for the dimension of space. Lengths are expressed in terms of $\bar{\ell}$. Importantly, we define the ratio of bending to stretching stiffnesses as $\kappa\!\equiv\!k_\theta/k_r\bar{\ell}^2$. For all calculations presented in what follows, we average the observables over about 100 independently created configurations. In what follows, the number of nodes $N$ employed in our networks is stated in the introduction of each data set.

\begin{figure}[ht!]
\includegraphics[width = 0.48\textwidth]{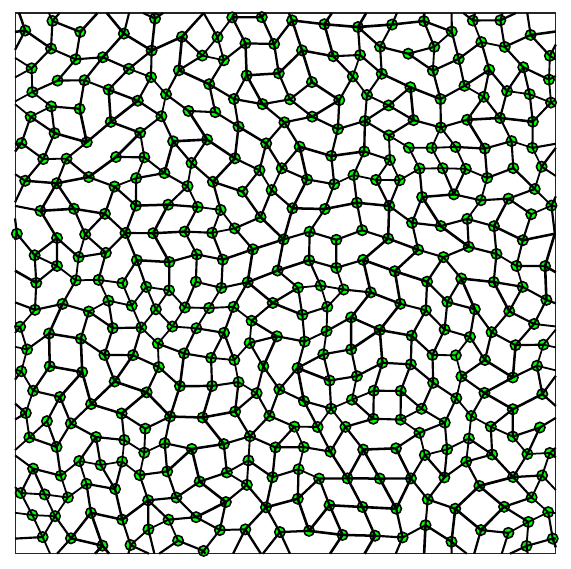}
  \caption{\footnotesize An example of an isostatic (i.e.~featuring $z\!=\!z_{\rm c}$), packing-derived network used in this study. Edges are diluted to reach the target coordination using the algorithm described in Ref.~\cite{phonon_width_2}, see text for discussion.}
  \label{fig:fig2}
\end{figure}

Our initial graphs were obtained by adopting the contact network of highly compressed soft-disc packings. To obtain networks at a target coordination $z$, we dilute the edges following the algorithm described in Ref.~\cite{phonon_width_2}; this algorithm produces networks with small node-to-node coordination fluctuations, and by such avoids the creation of locally rigid clusters~\cite{Thorpe_pre_1996,ellenbroek_rigidity_prl_2015}. An example of a network generated with this algorithm, at $z\!=\!z_{\rm c}$, is shown in Fig.~\ref{fig:fig2}.

We also carried out calculations in a three-dimensional complex elastic network model. In this model the disordered network considered are also derived from contact-networks of soft-sphere packings; however, in order to dilute the edges to reach some target coordination $z$, we employ the algorithm described in Ref.~\cite{anomalous_elasticity_soft_matter_2023}, which --- similarly to the algorithm employed for our 2D networks --- suppresses large node-to-node coordination fluctuations. In our 3D model we do not consider soft angular interactions; instead, for the sake of simplicity we follow Refs.~\cite{mw_maha_prl_2008,robbie_pre_2018} and connect soft springs of stiffness $\kappa k_r$ to all nearby nodes --- at distance $r\!\le\!1.1\bar{\ell}$ apart --- that are \emph{not} connected by a stiff (i.e.~of stiffness $k_r$) spring, and recall that $\kappa\!\ll\!1$. For this system, the coordination $z$ reported pertains to the stiff subnetwork. As we will show below, and as also demonstrated in~\cite{robbie_pre_2018}, these soft interactions give rise to the same scaling behavior as seen for our 2D networks with soft angular interactions. In Appendix~\ref{sec:appendix_angular} we show explicitly that the scaling properties of networks with additional soft radial springs, or with soft angular springs, is the same.

\begin{figure*}[!ht]
 \includegraphics[width = 1\textwidth]{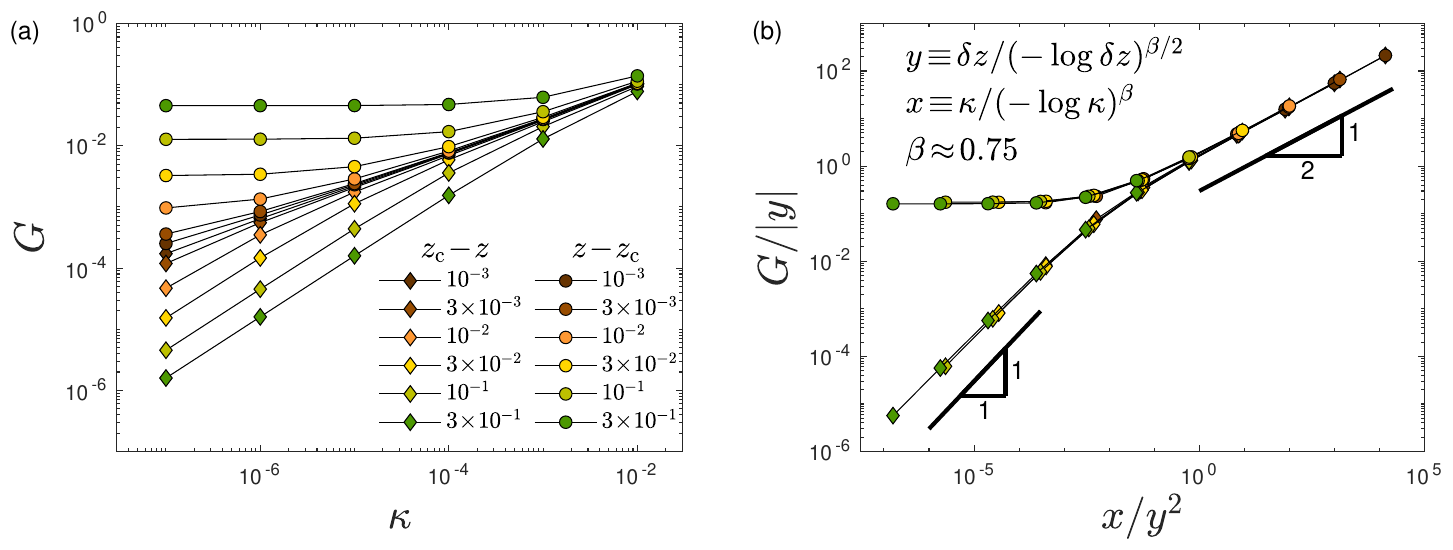}
  \caption{\footnotesize (a) The shear modulus $G$ of 2D isotropic, disordered elastic networks with $N\!=\!102400$ nodes, for various coordinations $z$ (see legend, and recall that $z_{\rm c}\!\equiv\!2\dbar$ with $\dbar$ denoting the spatial dimension), plotted against the stiffness ratio $\kappa$. (b) Same data as panel (a), plotted while rescaling the axes as described by the axes labels, see text for discussion.}
  \label{fig:fig3}
\end{figure*}

Finally, we also investigated the behavior of strain-stiffened networks in 2D; to this aim, we employed the following \emph{two-step procedure}~\cite{strain_stiffening_2023}: isotropic floppy networks with $z\!<\!z_{\rm c}$ were sheared --- using the standard athermal, quasistatic scheme~\cite{lemaitre2004_avalanches} --- while setting $\kappa\!=\!0$, until the ratio of the typical net-force on nodes to the typical compressive/tensile spring forces first drops below $10^{-8}$. The shear strain $\gamma_{\rm c}$ at which this happens is resolved up to strain increments of $10^{-6}$. Then, we gradually introduce angular interactions of dimensionless stiffness $\kappa$ -- in the form of the 2nd term of Eq.~(\ref{eq:potential_energy}) above. Once the soft interactions are introduced, the system is returned to a mechanically stable state by means of a potential energy minimization~\cite{fire}. Importantly, in this step the rest-angles considered are those defined in the \emph{isotropic, undeformed} network. This two-step procedure allows us to accurately determine the critical stiffening strain $\gamma_{\rm c}$, which is important since observables such as the shear modulus vary rapidly with strain at small $\kappa$ near $\gamma_{\rm c}$~\cite{strain_stiffening_2023}.

We conclude this Section with precise definitions of the key observables considered in this work. In Appendix~\ref{sec:appendix_angular_expressions} we provide easily implementable microscopic expressions for the angular potential and its derivatives with respect to coordinates and simple-shear-strain.

Our prime focus in this work is on the athermal shear modulus, defined as
\begin{equation}
G = \frac{1}{V}\bigg( \frac{\partial^2U}{\partial\gamma^2} - \frac{\partial^2U}{\partial\gamma\partial\xv}\cdot\calBold{H}^{-1}\cdot\frac{\partial^2U}{\partial\gamma\partial\xv}\bigg)\,,
\end{equation}
where $V\!=\!L^\dbar$ is the system's volume, $U$ is the potential energy, $\gamma$ is a simple shear strain parameter, $\xv$ denotes the nodes' coordinates, and $\calBold{H}\!\equiv\!\frac{\partial^2U}{\partial\xv\partial\xv}$ is the Hessian matrix. The shear modulus characterizes the macroscopic elastic response of the material; a microscopic characterization of the material's response is encoded in the typical nonaffine displacements squared, defined as
\begin{equation}\label{eq:nav_definition}
u_{\mbox{\scriptsize na}}^2 \equiv \frac{\uv_{\mbox{\scriptsize na}}\cdot\uv_{\mbox{\scriptsize na}}}{N} = \frac{\frac{\partial^2U}{\partial\gamma\partial\xv}\cdot\calBold{H}^{-2}\cdot\frac{\partial^2U}{\partial\gamma\partial\xv}}{N}\,.
\end{equation}
Finally, we study the spatial properties of the typical displacement response $\uv_{d}$ to local force dipoles~\cite{breakdown,sss_epje_2018,anomalous_elasticity_soft_matter_2023}, defined as
\begin{equation}\label{eq:dipole_response}
    \uv_d = \calBold{H}^{-1}\cdot\dv\,,
\end{equation}
where $\dv\!\equiv\!\frac{\partial r}{\partial \xv}$ is a unit force dipole applied to a radial spring with length $r$.

\section{Elasticity of Isotropic disordered networks}
\label{sec:results}

\subsection{Shear modulus of isotropic networks}

In Fig.~\ref{fig:fig3}a we present our results concerning the shear modulus $G(z,\kappa)$ and its dependence on coordination $z$ and the stiffness ratio $\kappa$ in our isotropic (undeformed) 2D disordered networks prepared as explained in Sect.~\ref{sec:model_and_methods}. Here we employed networks of $N\!=\!102400$ nodes at various coordinations $z$ as displayed in the figure legend. In Fig.~\ref{fig:fig3}b we show the same data rescaled by $y\!\equiv\!\delta z/(-\log\delta z)^{\beta/2}$ with $\beta\!=\!0.75$, and plotted against the scaling variable $x/y^2$ with $x\!\equiv\!\kappa/(-\log\kappa)^\beta$, to find a convincing data collapse. Our data suggest the scaling form
\begin{equation}\label{eq:shear_modulus_scaling_function}
    G(z,\kappa) = |\delta z|{\cal F}_{\pm}(\kappa/\delta z^2)\,,
\end{equation}
where ${\cal F}_\pm(\zeta)$ stands for a pair of scaling functions corresponding to $\delta z\!\gtrless\!0$, with the following properties
\begin{equation}\label{eq:scaling_function_properties}
\begin{matrix}
    {\cal F}_+(\zeta)   \sim  & \mbox{const}\,, & \zeta\ll 1\,,  \\
    {\cal F}_-(\zeta) \sim & \zeta\,, & \zeta \ll 1\,,  \\
    {\cal F}_\pm(\zeta) \sim & \sqrt{\zeta}\,, & \zeta \gg 1\,,
\end{matrix}
\end{equation}
and we deliberately suppress the logarithmic corrections, shown below to be a 2D effect. 

We next rationalize the observed scaling behavior using scaling argument and building on previously established results.

\begin{figure*}[btp]
 \includegraphics[width = 1\textwidth]{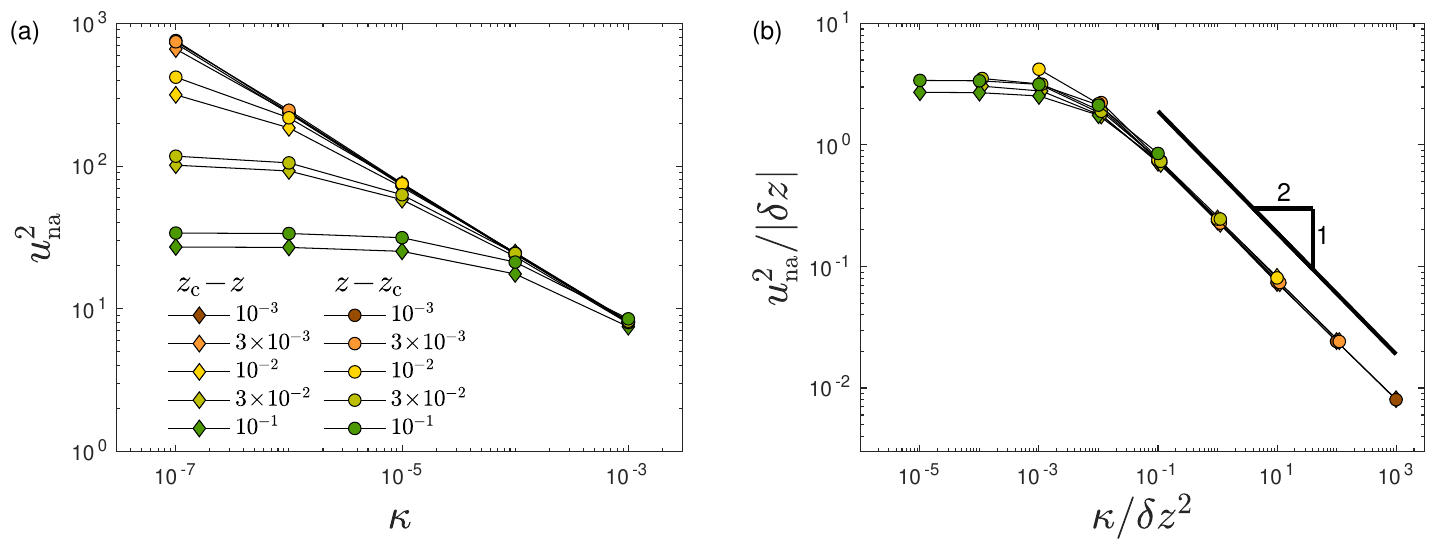}
  \caption{\footnotesize (a) The typical magnitude squared of nonaffine displacements (see Eq.~(\ref{eq:nav_definition}) for precise definition), calculated for 2D isotropic disordered networks at various coordinations $z$ as specified by the figure legend, and plotted against the stiffness ratio $\kappa$ (b) Same data as panel (a), but cast into a scaling form, see text for details and discussion.}
  \label{fig:fig4}
\end{figure*}

\vspace{-0.3cm}

\subsubsection{$G$ in the hypostatic regime $z\!<\!z_{\rm c}$}

It is known that for hypostatic ($z\!<\!z_{\rm c}$), isotropic, disordered spring networks in the absence of bending interactions, namely $\kappa\!=\!0$, the shear modulus is identically zero. 
In the limit $\kappa\!\to\!0$ the only (dimensionless) stiffness scale is $\kappa$, and therefore $G\!\sim\!\kappa$ in the $\kappa\!\to\!0$ limit, as indeed shown in previous work~\cite{chase_nature_physics_2011,mackintosh_pre_2016,Rens_JPCB_2016,robbie_nature_physics_2016,mackintosh_prl_2019}. The same conclusion can be obtain with more rigor following the framework of Ref.~\cite{robbie_pre_2018}, where the $\kappa\!\to\!0^+$ limit is considered, reducing the problem to a geometric one. In the same work and also earlier in Ref.~\cite{mw_maha_prl_2008} it was shown and rationalized that $G\!\sim\!1/\delta z$ for hypostatic elastic networks stabilized by weak interactions. Therefore, in the small-$\kappa$ limit, we expect that $G\!\sim\!\kappa/\delta z$, as indeed shown in Fig.~\ref{fig:fig3}, neglecting once again logarithmic corrections.

What would happen at larger $\kappa$? Since $\kappa$ is a stiffness scale, it is reasonable to compare it to the \emph{square} of the characteristic frequency $\omega_\star\!\sim\!\delta z$~\cite{matthieu_thesis,matthieu_PRE_2005,mw_EMT_epl,phonon_gap_2012,new_variational_argument_epl_2016} of the stiff subnetwork. We would therefore expect that once $\kappa$ is of order $\delta z^2$, the scaling behavior of $G$ should change. In other words, we expect that $\kappa/\omega_\star^2\!\sim\!\kappa/\delta z^2$ should form the relevant scaling variable (ignoring logarithmic corrections here and throughout what follows), as indeed seen in Fig.~\ref{fig:fig3}b and in the rest of the datasets presented in this work.

\vspace{-0.59cm}

\subsubsection{$G$ in the hyperstatic regime $z\!>\!z_{\rm c}$ }
\vspace{-0.3cm}
Since $\omega_\star\!\sim\!|\delta z|$ both above and below the isostatic point $z_{\rm c}$~\cite{matthieu_thesis,phonon_gap_2012}, the same argument for the relevant scaling variable $\kappa/\delta z^2$ should hold in the hyperstatic regime $z\!>\!z_{\rm c}$, as indeed observed in Fig.~\ref{fig:fig3}b. In addition, in the limit $\kappa\!\ll\!\delta z^2$ for $z\!>\!z_{\rm c}$ we expect the known behavior $G\!\sim\!\delta z$ to prevail, i.e.~$G$ should become independent of $\kappa$, as indeed observed in Fig.~\ref{fig:fig3}.

\subsubsection{$G$ in the critical regime $\kappa\!>\!\delta z^2$}

Interestingly, in the critical regime we find ${\cal F}_\pm(\zeta)\!\sim\!\sqrt{\zeta}$ (for $\zeta\!\gg\!1$). How can this scaling be rationalized? We expect that in the $\kappa\!\gg\!\delta z^2$ regime the stiff subnetwork's floppy modes, of frequency $\omega_\star\!\sim\!\delta z$ are stiffened by the angular interactions to feature frequencies of order $\sqrt{\kappa}$. We thus expect the characteristic frequency $\varpi$ of soft, nonphononic modes to follow
\begin{equation}\label{eq:nonphononic_freq}
    \varpi \sim \left\{ 
    \begin{matrix}
    \omega_\star \sim |\delta z| & \mbox{for} & \kappa \ll \delta z  \\
    \sqrt{\kappa} & \mbox{for} &\kappa\gg\delta z^2
    \end{matrix}\right. \,,
\end{equation}
ignoring logarithmic corrections.
Since the general relation between the shear modulus $G$ and the frequency of soft, nonphononic modes is theoretically predicted to follow $G\!\sim\!\omega_\star$ for $z\!>\!z_{\rm c}$, and in the absence of bending interactions, one could expect that, also in the critical $\kappa\!\gg\!\delta z^2$ regime, we would find $G\!\sim\!\varpi\!\sim\!\sqrt{\kappa}$. In practice, we find logarithmic corrections spelled out above -- cf.~Fig.~\ref{fig:fig3}b. In Sect.~\ref{sec:3D} below we show that these logarithmic corrections are absent in 3D.

\begin{figure*}[ht!]
 \includegraphics[width = 1\textwidth]{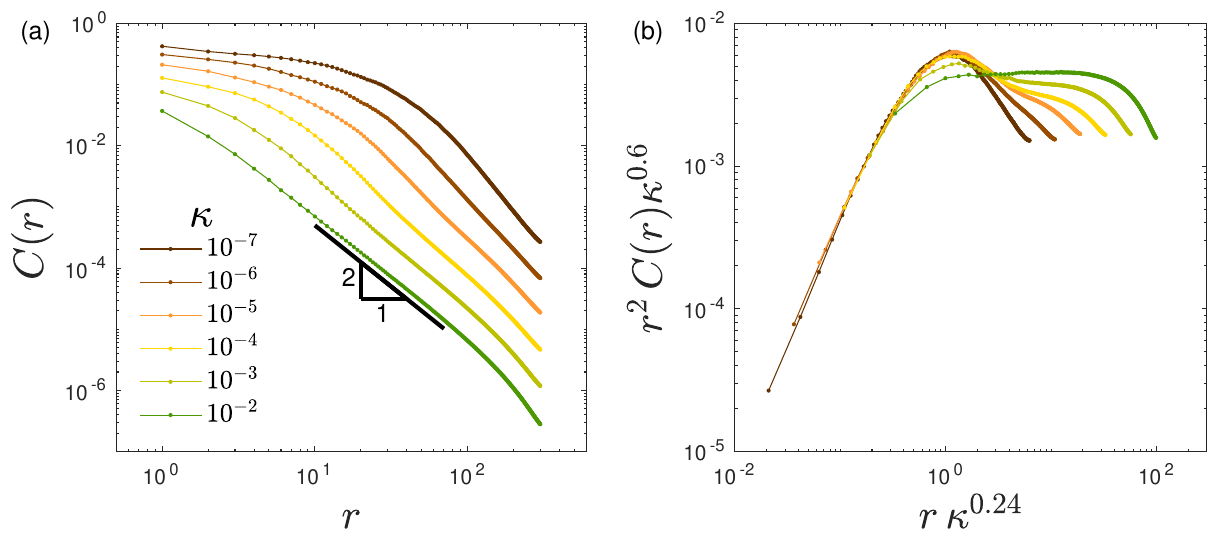}
  \caption{\footnotesize (a) Spatial decay of the displacement response functions $C(r)$ to local force dipoles, see text for definition. (b) Rescaling the axes allows to extract the $\kappa$-dependence of the length $\xi\!\sim\!\kappa^{-0.24}$ that marks the crossover between a nonaffine, disorder-dominated near field, and an affine, continuum-linear-elastic-like response in the far field, see text for further details and discussion.}
  \label{fig:fig5}
\end{figure*}

\subsection{Nonaffine displacements of isotropic networks}

We next turn to studying the typical nonaffine displacements squared $\nav$ (see Eq.~(\ref{eq:nav_definition}) above for a precise definition) in our 2D networks as a function of the coordination $z$ and the stiffness ratio $\kappa$. Fig.~\ref{fig:fig4} displays our results; in panel (a) we show the raw data, while panel (b) presents a scaling collapse, where, similarly to the analysis and argumentation provided above for the shear modulus, we consider the scaling variable $\kappa/\delta z^2$ to control the behavior of $\nav$ for various $\kappa$ and $z$. Different from the shear modulus analysis, here the $y$-axis is factored by $|\delta z|$, rationalized next for the different scaling regimes.

\subsubsection{$u_{\rm na}^2$ outside of the critical regime $\kappa\!\ll\!\delta z^2$}\label{sec:nav_small_kappa}

In Ref.~\cite{mw_maha_prl_2008} it was argued and demonstrated that $\nav\!\sim\!1/\delta z$ in the hypostatic $z\!<\!z_{\rm c}$ regime, while in Ref.~\cite{wouter_prl_2006} the same was argued and demonstrated for the hyperstatic $z\!>\!z_{\rm c}$ regime. Furthermore, a dimensional analysis of the $\kappa\!\to\!0$ expression for $\uv_{\mbox{\scriptsize na}}$ of Ref.~\cite{robbie_pre_2018} (see Eq.~(16) therein) implies that $\nav$ becomes \emph{independent} of $\kappa$ in the $\kappa\!\to\!0^+$ limit. We therefore expect that $\nav\!\sim\!1/\delta z$ for all $z,\kappa$, as indeed validated by the scaling collapse of Fig.~\ref{fig:fig4}b.

\subsubsection{$u_{\rm na}^2$ in the critical regime $\kappa\!\gg\!\delta z^2$}
\label{sec:nav_large_kappa}

Let us repeat the same line of argumentation as invoked for the scaling of the shear modulus in the critical regime $\kappa\!\gg\!\delta z^2$; in the absence of any angular interactions, namely on the $\kappa\!=\!0$ line, one finds $\nav\!\sim\!1/\delta z\!\sim\!1/\omega_\star$. Generalizing this result to a situation where nonphononic modes are stiffened by angular interactions --- such that $\varpi\!\sim\!\sqrt{\kappa}$ (cf.~Eq.~(\ref{eq:nonphononic_freq}) above) ---, one may then expect $\nav\!\sim\!1/\varpi\!\sim\!1/\sqrt{\kappa}$ for $\kappa\!\gg\!\delta z^2$. In Fig.~\ref{fig:fig4} this expectation is validated by our numerical simulations, to find very good agreement.

\subsection{The crossover length $\xi$ of isotropic networks}
\label{sec:length}

In the absence of bending interactions (i.e.~setting $\kappa\!=\!0$), disordered elastic networks feature a correlation length $\xi\!\sim\!1/\sqrt{\delta z}$ that marks the crossover between a near-field, nonaffine, disorder-dominated response to local force perturbations, to a far-field, affine, continuum-linear-elastic-like response, as shown in several prior works~\cite{Silbert_prl_2005,breakdown,anomalous_elasticity_soft_matter_2023}. Here we probe the $\kappa$-dependence of this correlation length in 2D networks of $N\!=\!409600$ nodes, and at $z\!=\!z_{\rm c}\!=\!4$ such that $\xi\!\sim\!L$ in the absence of angular interactions.

\begin{figure*}[ht!]
 \includegraphics[width = 1\textwidth]{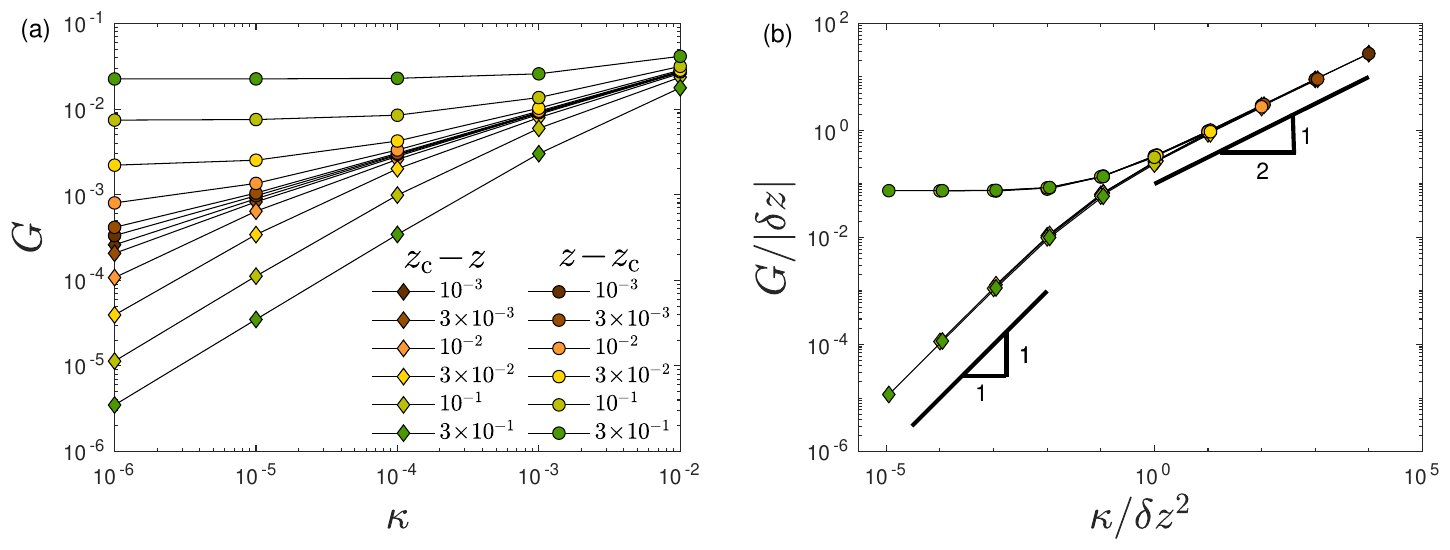}
  \caption{\footnotesize (a) The shear modulus of 3D elastic networks with coordinations various $z$ (see figure legend) --- stabilized by additional soft springs of dimensionless stiffness $\kappa\!\ll\!1$ (see text for precise details)---, is plotted against $\kappa$. (b) Same data as panel (a) cast into the predicted scaling form, see text for discussion.}
  \label{fig:fig6}
\end{figure*}

To extract the length $\xi(\kappa)$, we apply a unit dipole force to a single spring to obtain the displacement field $\uv_d$, as described in Eq.~(\ref{eq:dipole_response}). We then calculate the average of the displacement response squared --- denoted $C(r)$ --- over a shell of radius $r$ away from the force dipole. Continuum elasticity predict that the far field should decay as $1/r$ --- hence its squared amplitude as $1/r^2$ ---, as we indeed find for the largest $\kappa$'s probed, cf.~Fig.~\ref{fig:fig5}a.

How should $\xi$ scale with the stiffness ratio $\kappa$? Here we invoke the scaling argument put forward in Ref.~\cite{robbie_epje_2019}; a continuum, linear-elastic response can be seen if the system is large enough to accommodate elastic waves that are \emph{softer} than the characteristic frequency of nonphononic modes. This means that a crossover between the nonaffine, disordered response --- dominated by nonphononic modes --- to an affine, continuum-linear-elastic like response, dominated by elastic waves, occurs when
\begin{equation}
\frac{\sqrt{G}}{\xi} \sim \varpi\,.
\end{equation}
For our networks with $z\!=\!z_{\rm c}\!=\!4$ we always satisfy $\kappa\gg\delta z^2$ and therefore $\varpi\!\sim\!\sqrt{\kappa}$ (cf.~Eq.~(\ref{eq:nonphononic_freq})) and $G\!\sim\!\varpi\!\sim\!\sqrt{\kappa}$, resulting in the prediction $\xi\!\sim\!\kappa^{-1/4}$. Our numerical results are shown in Fig.~\ref{fig:fig5}b, where the rescaling of both axes leads to an alignment and collapse of the peaks. We find $\xi\!\sim\!\kappa^{-0.24}$, very close to the prediction $\xi\!\sim\!\kappa^{-1/4}$, supporting the scaling argument spelled out above. The slight disagreement with the predicted exponent of $-1/4$ could stem from the logarithmic corrections, not accounted for here.

\subsection{Elasticity of disordered networks in 3D}
\label{sec:3D}

In this final Subsection we show that the scaling behavior seen in 2D networks endowed with soft angular interactions is also seen in 3D, using a different, simpler form of soft interactions. In particular, instead of defining angular interactions in 3D, we simply connect nearby nodes --- that are not already connected by stiff springs --- by soft springs, of stiffness $\kappa k_r$, and see Sect.~\ref{sec:model_and_methods} for further details. In Appendix~\ref{sec:appendix_angular} we show that, in 2D, the scaling properties of networks endowed with weak angular or weak radial interactions -- are the same.

We focus here on the shear modulus $G(\kappa,z)$, similar to the analysis presented for our 2D networks in Fig.~\ref{fig:fig3}. The results are displayed in Fig.~\ref{fig:fig6}; panel (a) shows the raw data for the shear modulus for a variety of coordinations $z$ and stiffness ratios $\kappa$, while panel (b) employs a rescaling of the axes to find a data collapse. For our 3D networks, are data are perfectly consistent with the same scaling form given by Eqs.~(\ref{eq:shear_modulus_scaling_function}) and (\ref{eq:scaling_function_properties}) but without the logarithmic corrections, which appear to be a 2D effect. These results, together with the results presented in Appendix~\ref{sec:appendix_angular}, indicate that the precise functional form of the soft interactions does not change the scaling behavior observed. We note finally that the exponents obtained from our scaling analysis of 3D networks agree perfectly with Effective Medium Theory calculations in 2D~\cite{chase_nature_physics_2011,Xiaoming_emt_2013}.

\vspace{-0.4cm}

\section{Elasticity of strain-stiffened networks}
\label{sec:strain_stiffened}

Strain stiffening refers to the phenomenon in which a hypostatic (floppy) network with soft (but nonzero) angular (bending) interactions --- with $\kappa\!\ll\!\delta z^2$ --- is deformed such that, as some critical strain $\gamma_{\rm c}$ is approached, the shear modulus grows substantially, from $G/k_r\!\sim\!\kappa$ (for $\kappa\!\ll\!\delta z^2$) in the undeformed, isotropic states, to $G/k_r\!\sim\!{\cal O}(1)$ as $\gamma\!\to\!\gamma_{\rm c}$~\cite{robbie_thesis,wouter_pre_2017,merkel_pnas_2019,strain_stiffening_2023}. Here we study crossover effects between $\kappa\!\ll\!\delta z^2$ and $\kappa\!\gg\!\delta z^2$ in strain-stiffened networks in 2D, similar to our analysis presented in the previous Section for isotropic, disordered networks. 

In order to sharply identify the critical strain $\gamma_{\rm c}$, we employ the following `two-step' procedure~\cite{strain_stiffening_2023}: we first set $\kappa\!=\!0$ in our isotropic networks, and employ the athermal, quasistatic deformation scheme in which small strain increments are applied, following each one with a potential-energy minimization~\cite{fire}. Under $\kappa\!=\!0$ conditions, the shear modulus is identically zero for $\gamma\!<\!\gamma_{\rm c}$, and jumps discontinuously to $G\!\sim\!\delta z$ at $\gamma\!=\!\gamma_{\rm c}$. Details about the detection of $\gamma_{\rm c}$ are provided in Sect.~\ref{sec:model_and_methods} above.

Once strain-stiffened networks at $\gamma\!=\!\gamma_{\rm c}$ are at hand, we then gradually introduce the soft angular interactions, similarly to the athermal quasistatic deformation scheme: after increasing $\kappa$ to some target value, we restore mechanical equilibrium by a potential energy minimization. We then analyze the elastic properties of these networks as a function of $\kappa$ and $z\!<\!z_{\rm c}$, and we note that, for these calculations we employed networks of $N\!=\!25600$ nodes. 

\begin{figure}[ht!]
 \includegraphics[width = 0.5\textwidth]{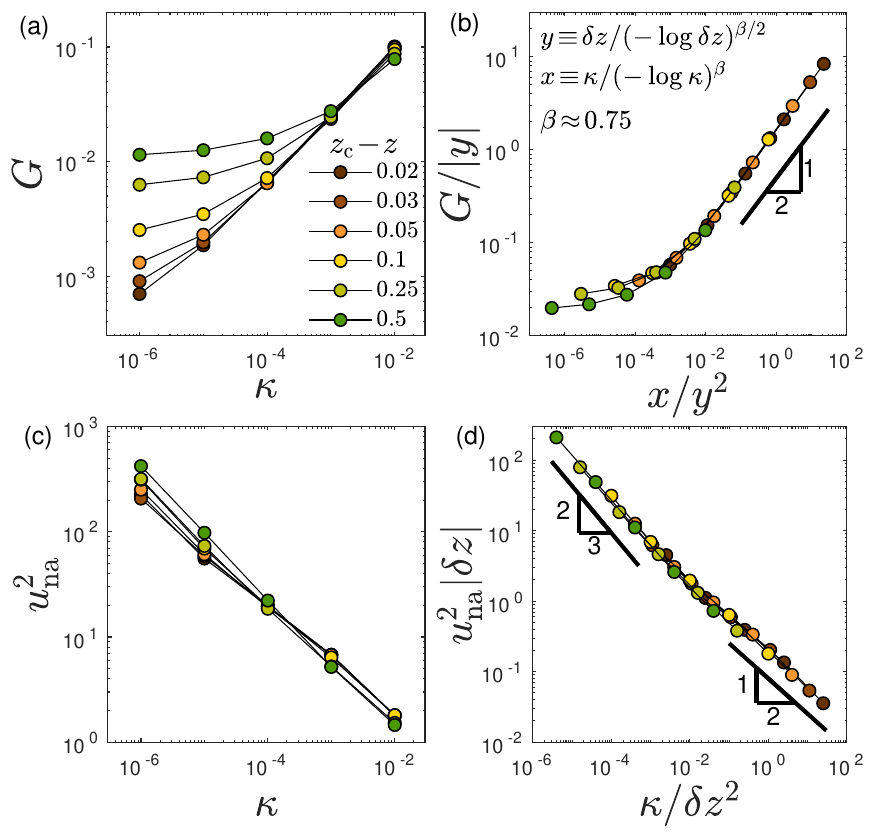}
  \caption{\footnotesize (a) The shear modulus $G$ of 2D hypostatic strain-stiffened networks at various coordinations as indicated by the legend (b) Same data as panel (a), cast into the expected scaling form, see text for details. (c) The characteristic magnitude squared of nonaffine displacements, $\nav$, for the same coordinations as indicated in the legend of panel (a). (d) Scaling collapse of $\nav$ according to the expected form, see text for detailed discussion.}
  \label{fig:fig7}
\end{figure}

The results are presented in Fig.~\ref{fig:fig7}; as expected, the relevant scaling variable is again $\kappa/\delta z^2$ (neglecting logarithmic corrections), just like we have established for isotropic networks above. In panel (a) we show the raw data for the shear modulus $G$ and the typical square of the node-wise nonaffine displacements $\nav$. Despite that here $z\!<\!z_{\rm c}$, the shear modulus data precisely resembles the shear modulus data of isotropic, $z\!>\!z_{\rm c}$ states, cf.~Fig.~\ref{fig:fig3}a, and follows the \emph{exact} same scaling form of ${\cal F}_+(\zeta)$ (cf.~Eq.~(\ref{eq:shear_modulus_scaling_function})), namely $G\!\sim\!|\delta z|$ (i.e.~independent of $\kappa$, and notice that here $\delta z\!<\!0$) for $\kappa\!\ll\!\delta z^2$, and $G\!\sim\!\sqrt{\kappa}$ (independent of $\delta z$) for $\kappa\!\gg\!\delta z^2$. These scaling behaviors are validated by the scaling collapse of Fig.~\ref{fig:fig7}b.

In Fig.~\ref{fig:fig7}b we show data for the typical nonaffine displacements squared $\nav$ of strain-stiffened networks. In the large-$\kappa$ regime, we find the same behavior as seen for isotropic networks, namely that $\nav\!\sim\!1/\sqrt{\kappa}$ (see Fig.~\ref{fig:fig7}d), independent of $\delta z$ and consistent with expectations as discussed in Sect.~\ref{sec:nav_large_kappa} above for isotropic networks. Interestingly, and in stark contrast with the large-$\kappa$ regime, the scaling behavior of $\nav$ with $\kappa$ is \emph{different} compared to that seen for isotropic networks in the $\kappa\!\ll\!\delta z^2$ regime; in particular, in this regime we find a \emph{singular} scaling $\nav\!\sim\!\kappa^{-2/3}$ as predicted theoretically in Ref.~\cite{strain_stiffening_2023} for strain-stiffened states, and observed in previous simulational work~\cite{mackintosh_prl_2019,fred_prl_2023} for networks with various geometries. This should be compared to the behavior of $\nav$ in the small-$\kappa$ regime of isotropic states, where it is predicted and observed to become $\kappa$-independent (for any $z$), cf.~discussion in Sect.~\ref{sec:nav_small_kappa}. The reason for this difference is that strain-stiffened networks with $\kappa\!\ll\!\delta z^2$ feature soft modes of frequency $\omega_\kappa\sim\!\kappa^{1/3}$~\cite{strain_stiffening_2023}, and one can show that $\nav\!\sim\!\omega_\kappa^{-2}$. This situation is very similar to that seen in floppy networks of rigid struts sheared with overdamped dynamics~\cite{gustavo_pre_2014}, where $\nav\!\sim\!\omega_{\mbox{\tiny min}}^{-2}$ where  $\omega_{\mbox{\tiny min}}$ (in the notations of~\cite{gustavo_pre_2014}) represents the dimensionless force unbalance the nodes experience under shear.

\section{summary and discussion}
\label{sec:summary}

In this work we explored the elastic properties of disordered Hookean spring networks featuring angular interactions whose characteristic stiffness is far smaller than the stiffness associated with the networks' radial springs. We traced out the different scaling regimes in terms of the coordination $z$ and the ratio $\kappa$ between the angular- and radial-spring-stiffnesses. This was done in 2D both for undeformed, isotropic systems and for strain-stiffened states, the latter is restricted (by construction) to hypostatic ($z\!<\!z_{\rm c}$) systems. In 3D we explored similar questions, but substituted the angular interactions employed in 2D with a simpler form of weak interactions. 

We found and rationalized that the scaling variable that controls the elastic properties of these complex networks is the ratio $\kappa/\delta z^2$. This ratio expresses the competition between the characteristic stiffness $\sim\!\delta z^2$ of nonphononic soft modes of the stiff subnetwork (i.e.~in the absence of angular interactions), and the stiffness $\sim\!\kappa$ of angular interactions. 

Interestingly, we found that in the critical $\kappa\!\gg\!\delta z^2$ regime, $G$ scales as $\sqrt{\kappa}$ --- with logarithmic corrections in 2D --- which echoes the theoretically predicted dependence of the shear modulus on the characteristic frequency of nonphononic soft modes in simple hyperstatic ($z\!>\!z_{\rm c}$) elastic networks, namely $G\!\sim\!\omega_\star\!\sim\!\delta z$. The same behavior, namely that $G\!\sim\!\sqrt{\kappa}$, is also seen for hypostatic, strain-stiffened networks in the $\kappa\!\gg\!\delta z^2$ regime. Our results for isotropic networks agree --- exactly in 3D, and with logarithmic corrections in 2D --- with Effective Medium Theory calculations in 2D~\cite{chase_nature_physics_2011,Xiaoming_emt_2013}. In Ref.~\cite{chase_nature_physics_2011}, it is observed that $G\!\sim\!\kappa^{\chi}$ in the large-$\kappa$ regime of 2D randomly diluted triangular lattices, with $\chi\!=\!0.46\pm0.07$, not far from the $\sim\!\sqrt{\kappa/(-\log\kappa)^{0.75}}$ scaling observed in our 2D disordered networks. 

We additionally investigated how the lengthscale $\xi$ that marks the crossover between a near-field, disorder dominated response to local force dipoles, and a far-field, continuum-linear-elastic-like response, depends on $\kappa$ in isostatic, isotropic networks. Our scaling arguments predict that $\xi\!\sim\!\kappa^{-1/4}$ in 2D, for which we find satisfying agreement with our numerical results. 

\begin{figure*}[!ht]
 \includegraphics[width = 1\textwidth]{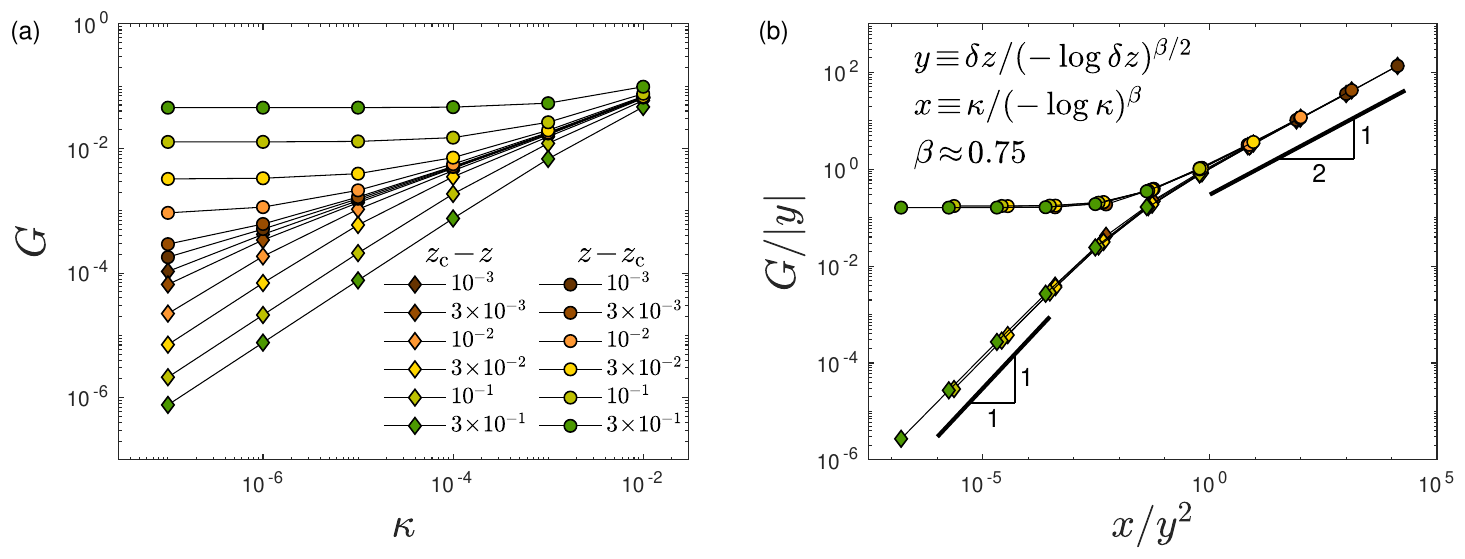}
  \caption{\footnotesize Same as Fig.~\ref{fig:fig3}, calculated for networks in which, instead of considering angular interactions, we connect radial springs of stiffness $\kappa k_r$ between closeby pairs of nodes that are not already connected by a stiff spring, as done in 3D above (see Fig.~\ref{fig:fig6}). We find the same scaling behavior observed in systems with angular interactions, supporting that the precise functional form of the weak interactions does not alter the mechanical scaling properties of these complex networks.}
  \label{fig:weak_springs_2D}
\end{figure*}

Finally, we pointed out some interesting differences in the elastic properties of strain-stiffened networks compared to isotropic ones. In particular, we find that the typical nonaffine displacement squared, $\nav$, is regular in $\kappa$ (for small $\kappa/\delta z^2$) in isotropic networks, but singular $\sim\!\kappa^{-2/3}$ in strain-stiffened networks, in agreement with earlier theoretical predictions~\cite{mw_maha_prl_2008,robbie_pre_2018,strain_stiffening_2023} and computational work~\cite{mackintosh_prl_2019,fred_prl_2023}, and similar to observations in sheared floppy networks of rigid struts~\cite{gustavo_pre_2014}. Moreover, the scaling collapse of $\nav$ presented in Fig.~\ref{fig:fig7}d suggests that, for $\kappa\!\ll\!\delta z^2$, $\nav\!\sim\!\delta z^{1/3}\kappa^{-2/3}$. This scaling with $\delta z$ lacks a theoretical explanation, and is left to be resolved in future studies. 

\acknowledgements
We thank Eran Bouchbinder,  and Gustavo D\"uring and Chase Broedersz for invaluable discussions.

\appendix

\section{Equivalence of weak angular interactions and weak radial interactions}
\label{sec:appendix_angular}

In this Appendix we show that replacing weak angular interactions with additional weak radial springs gives rise to the same scaling behavior of elastic properties. We consider the same networks as in the main text, but this time, instead of introducing a soft potential that depends on the angles as described in Fig.~\ref{fig:theta_illustration} below, we instead connect weak springs of stiffness $\kappa\,k_{r}$ to pairs of nearby nodes that are \emph{not} already connected by a regular spring. We measured the shear modulus $G$ of these complex networks, and report the results in Fig.~\ref{fig:weak_springs_2D}. We find the same scaling behavior as reported in Fig.~\ref{fig:fig3} for networks endowed with angular interactions.

\begin{figure}[ht!]
 \includegraphics[width = 0.25\textwidth]{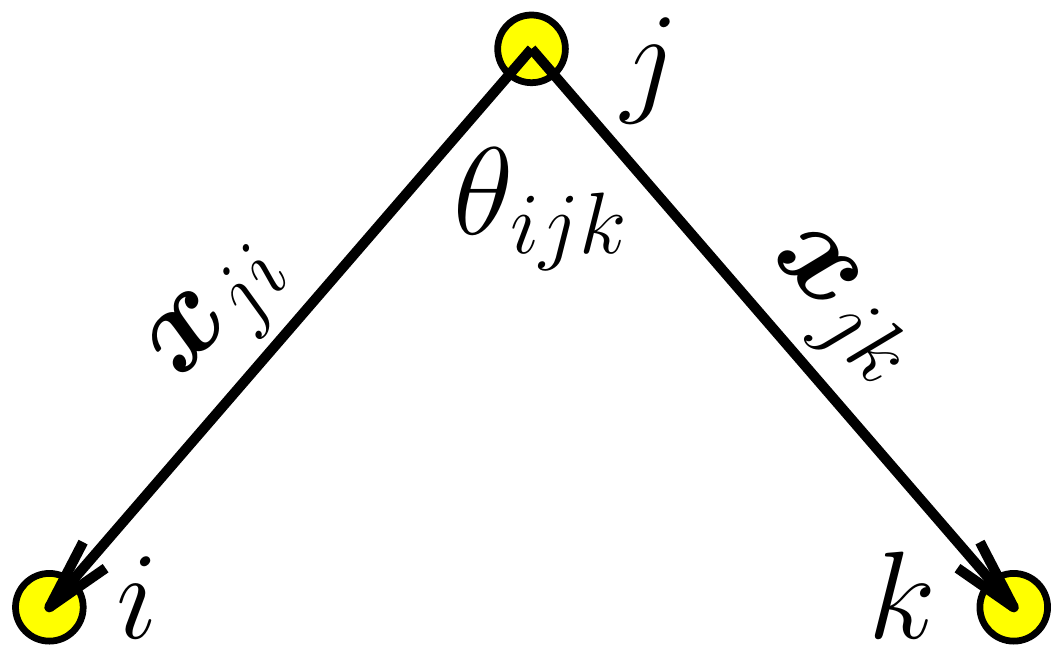}
  \caption{\footnotesize The convention chosen here for angles $\theta_{ijk}$ in 2D.}
  \label{fig:theta_illustration}
\end{figure}

\section{Angular interactions}
\label{sec:appendix_angular_expressions}

In this Appendix we provide easily implementable expressions for the bending interactions. We consider a potential energy of the form
\[
U_{\mbox{\tiny bend}}(\xv) = \frac{1}{2}\!\!\!\!\! \sum_{\mbox{\tiny angles }ijk}\!\!\!\!\!(\theta_{ijk}-\theta^{(0)}_{ijk})^2\,,
\]
where the sum runs over relevant angles $\theta_{ijk}$, and $\theta^{(0)}$ denote the rest-angles. We employ a convention in which $j$ is the index of the shared node forming the angle $\theta_{ijk}$, see Fig.~\ref{fig:theta_illustration}; with this construction, one has $\frac{\partial\theta_{ijk}}{\partial\xv_j}\!=\!-\frac{\partial\theta_{ijk}}{\partial\xv_i}\!-\!\frac{\partial\theta_{ijk}}{\partial\xv_k}$ and that the latter two terms assume the form
\begin{eqnarray}
\frac{\partial\theta_{ijk}}{\partial\xv_i} & = &  \frac{\xv_{ji}}{r_{ji}^2}\cdot(\hat{\yv}\hat{\xv}-\hat{\xv}\hat{\yv})\,, \nonumber\\
\frac{\partial\theta_{ijk}}{\partial\xv_k} & = & \frac{\xv_{jk}}{r_{jk}^2}\cdot(\hat{\xv}\hat{\yv}-\hat{\yv}\hat{\xv})\,,\nonumber 
\end{eqnarray} 
where $\xv_{ji}\!\equiv\!\xv_i\!-\!\xv_j$, $r_{ji}\!\equiv\!\sqrt{\xv_{ji}\!\cdot\!\xv_{ji}}$, and $\hat{\xv},\hat{\yv}$ denote the Cartesian unit vectors. The second derivatives of angles with respect to coordinates read
\begin{eqnarray}
    \frac{\partial^2\theta_{ijk}}{\partial\xv_i\partial\xv_i} & = & \bigg(\frac{\calBold{I}}{r_{ji}^2}-2\frac{\xv_{ji}\xv_{ji}}{r_{ji}^4}\bigg)\cdot(\hat{\yv}\hat{\xv}\!-\!\hat{\xv}\hat{\yv})\,, \nonumber \\ 
    & = & \frac{1}{r_{ji}^2}\bigg(
    \begin{matrix}
-2(\nv_{ji}\!\cdot\!\hat{\yv})(\nv_{ji}\!\cdot\!\hat{\xv}) & 2(\nv_{ji}\!\cdot\!\hat{\xv})^2\!-\!1\\1\!-\!2(\nv_{ji}\!\cdot\!\hat{\yv})^2 & 2(\nv_{ji}\!\cdot\!\hat{\yv})(\nv_{ji}\!\cdot\!\hat{\xv})
\end{matrix}\bigg)\,, \nonumber \\
 \frac{\partial^2\theta_{ijk}}{\partial\xv_k\partial\xv_k} & = & \bigg(\frac{\calBold{I}}{r_{jk}^2}-2\frac{\xv_{jk}\xv_{jk}}{r_{jk}^4}\bigg)\cdot(\hat{\xv}\hat{\yv}-\hat{\yv}\hat{\xv}) \nonumber \\
& = &  \frac{1}{r_{jk}^2}\bigg(
    \begin{matrix}
2(\nv_{jk}\!\cdot\!\hat{\yv})(\nv_{jk}\!\cdot\!\hat{\xv}) & 1-2(\nv_{jk}\!\cdot\!\hat{\xv})^2\\2(\nv_{jk}\!\cdot\!\hat{\yv})^2-1 & -2(\nv_{jk}\!\cdot\!\hat{\yv})(\nv_{jk}\!\cdot\!\hat{\xv})
\end{matrix}\bigg)\,, \nonumber \\
    \frac{\partial^2\theta_{ijk}} {\partial\xv_j\partial\xv_i} & = &-\frac{\partial^2\theta_{ijk}}{\partial\xv_i\partial\xv_i}\,,\nonumber \\
    \frac{\partial^2\theta_{ijk}}{\partial\xv_j\partial\xv_k} & = &  -\frac{\partial^2\theta_{ijk}}{\partial\xv_k\partial\xv_k}\,. \nonumber\\
    \frac{\partial^2\theta_{ijk}}{\partial\xv_j\partial\xv_j} & = &   \frac{\partial^2\theta_{ijk}}{\partial\xv_i\partial\xv_i} + \frac{\partial^2\theta_{ijk}}{\partial\xv_k\partial\xv_k}\,, \nonumber
\end{eqnarray}
where $\nv_{jk}\!\equiv\!\xv_{jk}/r_{jk}$, and notice that the off-diagonal terms $\frac{\partial^2\theta_{ijk}}{\partial\xv_i\partial\xv_k}\!=\!0$. For ease of notations we define $\Phi(\theta)\!\equiv\!\frac{1}{2}(\theta-\theta^{(0)})^2$, $\Phi'\!\equiv\!\partial\Phi/\partial\theta$, etc., then
\begin{equation*}
    \frac{\partial U_{\mbox{\tiny bend}}}{\partial\xv} = \!\!\! \sum_{\mbox{\tiny angles }ijk}\!\!\! \Phi'(\theta_{ijk})\frac{\partial\theta_{ijk}}{\partial\xv}\,,
\end{equation*}
and
\begin{eqnarray}
\frac{\partial^2U_{\mbox{\tiny bend}}}{\partial\xv\partial\xv}& =& \!\!\! \sum_{\mbox{\tiny angles }ijk}\!\!\! \Phi''(\theta_{ijk})\frac{\partial\theta_{ijk}}{\partial\xv}\frac{\partial\theta_{ijk}}{\partial\xv} \nonumber \\
&&+ \!\!\! \sum_{\mbox{\tiny angles }ijk} \!\!\!\Phi'(\theta_{ijk})\frac{\partial^2\theta_{ijk}}{\partial\xv\partial\xv}\,. \nonumber
\end{eqnarray}

We next spell out derivatives with respect to shear strain; we consider the transformation of coordinates $\xv\!\to\!{\bm H}(\gamma)\!\cdot\!\xv$ where $\gamma$ is a shear-strain parameter, and
\[
    {\bm H}(\gamma) = \left( \begin{array}{cc}1&\gamma\\0&1\end{array}\right)\,,
\]
then
\begin{eqnarray}
     \frac{\partial\theta_{ijk}}{\partial\gamma} & = & (\nv_{ji}\cdot\hat{\yv})^2 - (\nv_{jk}\cdot\hat{\yv})^2\,, \nonumber \\
     \frac{\partial^2\theta_{ijk}}{\partial\gamma^2} & = & 
     2(\nv_{jk}\cdot\hat{\yv})^3(\nv_{jk}\cdot\hat{\xv}) - 2(\nv_{ji}\cdot\hat{\yv})^3(\nv_{ji}\cdot\hat{\xv})\,. \nonumber
\end{eqnarray}
With these relations, one has
\begin{eqnarray}
\frac{\partial U_{\mbox{\tiny bend}}}{\partial\gamma} & = & \!\!\!\! \sum_{\mbox{\tiny angles }ijk}\!\!\!\!\! \Phi'(\theta_{ijk})\frac{\partial\theta_{ijk}}{\partial\gamma} \,, \nonumber \\
\frac{\partial^2U_{\mbox{\tiny bend}}}{\partial\gamma^2} & = &\!\!\!\!\! \sum_{\mbox{\tiny angles }ijk}\!\!\!\!\! \Phi''(\theta_{ijk})\bigg(\frac{\partial\theta_{ijk}}{\partial\gamma}\bigg)^2 \!\!+\!\! \!\!\!\! \sum_{\mbox{\tiny angles }ijk} \!\!\!\!\!\!\Phi'(\theta_{ijk})\frac{\partial^2\theta_{ijk}}{\partial\gamma^2}\,, \nonumber
\end{eqnarray}

Finally the mixed derivatives of the angles read
\begin{eqnarray}
\frac{\partial^2\theta_{ijk}}{\partial\gamma\partial\xv_i} & = &\frac{\nv_{ji}\!\cdot\!\hat{\yv}}{r_{ji}}\big( 2(\nv_{ji}\!\cdot\!\hat{\xv})^2-1\big)\hat{\yv}-2\frac{(\nv_{ji}\cdot\hat{\xv})(\nv_{ji}\!\cdot\!\hat{\yv})^2}{r_{ji}}\hat{\xv}\,, \nonumber \\
\frac{\partial^2\theta_{ijk}}{\partial\gamma\partial\xv_k} & = & 2\frac{(\nv_{jk}\!\cdot\!\hat{\xv})(\nv_{jk}\!\cdot\!\hat{\yv})^2}{r_{jk}}\hat{\xv} -  \frac{\nv_{jk}\!\cdot\!\hat{\yv}}{r_{jk}}\big( 2(\nv_{jk}\!\cdot\!\hat{\xv})^2-1\big)\hat{\yv}\,, \nonumber \\
\frac{\partial^2\theta_{ijk}}{\partial\gamma\partial\xv_j} & = & - \frac{\partial^2\theta_{ijk}}{\partial\gamma\partial\xv_i} - \frac{\partial^2\theta_{ijk}}{\partial\gamma\partial\xv_k}\,, \nonumber
\end{eqnarray}
and then
\[
\frac{\partial^2U_{\mbox{\tiny bend}}}{\partial\gamma\partial\xv} =   \!\!\!\! \sum_{\mbox{\tiny angles }ijk} \!\!\!\!\!\!\Phi''(\theta_{ijk})\frac{\partial\theta_{ijk}}{\partial\gamma}\frac{\partial\theta_{ijk}}{\partial\xv} + \!\!\!\! \sum_{\mbox{\tiny angles }ijk} \!\!\!\!\!\!\Phi'(\theta_{ijk})\frac{\partial^2\theta_{ijk}}{\partial\gamma\partial\xv}\,.
\]


\begin{thebibliography}{41}%
\makeatletter
\providecommand \@ifxundefined [1]{%
 \@ifx{#1\undefined}
}%
\providecommand \@ifnum [1]{%
 \ifnum #1\expandafter \@firstoftwo
 \else \expandafter \@secondoftwo
 \fi
}%
\providecommand \@ifx [1]{%
 \ifx #1\expandafter \@firstoftwo
 \else \expandafter \@secondoftwo
 \fi
}%
\providecommand \natexlab [1]{#1}%
\providecommand \enquote  [1]{``#1''}%
\providecommand \bibnamefont  [1]{#1}%
\providecommand \bibfnamefont [1]{#1}%
\providecommand \citenamefont [1]{#1}%
\providecommand \href@noop [0]{\@secondoftwo}%
\providecommand \href [0]{\begingroup \@sanitize@url \@href}%
\providecommand \@href[1]{\@@startlink{#1}\@@href}%
\providecommand \@@href[1]{\endgroup#1\@@endlink}%
\providecommand \@sanitize@url [0]{\catcode `\\12\catcode `\$12\catcode
  `\&12\catcode `\#12\catcode `\^12\catcode `\_12\catcode `\%12\relax}%
\providecommand \@@startlink[1]{}%
\providecommand \@@endlink[0]{}%
\providecommand \url  [0]{\begingroup\@sanitize@url \@url }%
\providecommand \@url [1]{\endgroup\@href {#1}{\urlprefix }}%
\providecommand \urlprefix  [0]{URL }%
\providecommand \Eprint [0]{\href }%
\providecommand \doibase [0]{https://doi.org/}%
\providecommand \selectlanguage [0]{\@gobble}%
\providecommand \bibinfo  [0]{\@secondoftwo}%
\providecommand \bibfield  [0]{\@secondoftwo}%
\providecommand \translation [1]{[#1]}%
\providecommand \BibitemOpen [0]{}%
\providecommand \bibitemStop [0]{}%
\providecommand \bibitemNoStop [0]{.\EOS\space}%
\providecommand \EOS [0]{\spacefactor3000\relax}%
\providecommand \BibitemShut  [1]{\csname bibitem#1\endcsname}%
\let\auto@bib@innerbib\@empty
\bibitem [{\citenamefont {Alexander}(1998)}]{shlomo}%
  \BibitemOpen
  \bibfield  {author} {\bibinfo {author} {\bibfnamefont {S.}~\bibnamefont
  {Alexander}},\ }\bibfield  {title} {\bibinfo {title} {Amorphous solids: their
  structure, lattice dynamics and elasticity},\ }\href
  {https://doi.org/http://dx.doi.org/10.1016/S0370-1573(97)00069-0} {\bibfield
  {journal} {\bibinfo  {journal} {Phys. Rep.}\ }\textbf {\bibinfo {volume}
  {296}},\ \bibinfo {pages} {65 } (\bibinfo {year} {1998})}\BibitemShut
  {NoStop}%
\bibitem [{\citenamefont {Wyart}(2005)}]{matthieu_thesis}%
  \BibitemOpen
  \bibfield  {author} {\bibinfo {author} {\bibfnamefont {M.}~\bibnamefont
  {Wyart}},\ }\bibfield  {title} {\bibinfo {title} {On the rigidity of
  amorphous solids},\ }\href {https://doi.org/10.1051/anphys:2006003}
  {\bibfield  {journal} {\bibinfo  {journal} {Ann. Phys. Fr.}\ }\textbf
  {\bibinfo {volume} {30}},\ \bibinfo {pages} {1} (\bibinfo {year}
  {2005})}\BibitemShut {NoStop}%
\bibitem [{\citenamefont {Wyart}\ \emph {et~al.}(2008)\citenamefont {Wyart},
  \citenamefont {Liang}, \citenamefont {Kabla},\ and\ \citenamefont
  {Mahadevan}}]{mw_maha_prl_2008}%
  \BibitemOpen
  \bibfield  {author} {\bibinfo {author} {\bibfnamefont {M.}~\bibnamefont
  {Wyart}}, \bibinfo {author} {\bibfnamefont {H.}~\bibnamefont {Liang}},
  \bibinfo {author} {\bibfnamefont {A.}~\bibnamefont {Kabla}},\ and\ \bibinfo
  {author} {\bibfnamefont {L.}~\bibnamefont {Mahadevan}},\ }\bibfield  {title}
  {\bibinfo {title} {Elasticity of floppy and stiff random networks},\ }\href
  {https://doi.org/10.1103/PhysRevLett.101.215501} {\bibfield  {journal}
  {\bibinfo  {journal} {Phys. Rev. Lett.}\ }\textbf {\bibinfo {volume} {101}},\
  \bibinfo {pages} {215501} (\bibinfo {year} {2008})}\BibitemShut {NoStop}%
\bibitem [{\citenamefont {Ellenbroek}\ \emph {et~al.}(2009)\citenamefont
  {Ellenbroek}, \citenamefont {Zeravcic}, \citenamefont {van Saarloos},\ and\
  \citenamefont {van Hecke}}]{Ellenbroek_2009}%
  \BibitemOpen
  \bibfield  {author} {\bibinfo {author} {\bibfnamefont {W.~G.}\ \bibnamefont
  {Ellenbroek}}, \bibinfo {author} {\bibfnamefont {Z.}~\bibnamefont
  {Zeravcic}}, \bibinfo {author} {\bibfnamefont {W.}~\bibnamefont {van
  Saarloos}},\ and\ \bibinfo {author} {\bibfnamefont {M.}~\bibnamefont {van
  Hecke}},\ }\bibfield  {title} {\bibinfo {title} {Non-affine response: Jammed
  packings vs. spring networks},\ }\href
  {https://doi.org/10.1209/0295-5075/87/34004} {\bibfield  {journal} {\bibinfo
  {journal} {Europhys. Lett.}\ }\textbf {\bibinfo {volume} {87}},\ \bibinfo
  {pages} {34004} (\bibinfo {year} {2009})}\BibitemShut {NoStop}%
\bibitem [{\citenamefont {Wyart}(2010)}]{mw_EMT_epl}%
  \BibitemOpen
  \bibfield  {author} {\bibinfo {author} {\bibfnamefont {M.}~\bibnamefont
  {Wyart}},\ }\bibfield  {title} {\bibinfo {title} {Scaling of phononic
  transport with connectivity in amorphous solids},\ }\href
  {http://stacks.iop.org/0295-5075/89/i=6/a=64001} {\bibfield  {journal}
  {\bibinfo  {journal} {Europhys. Lett.}\ }\textbf {\bibinfo {volume} {89}},\
  \bibinfo {pages} {64001} (\bibinfo {year} {2010})}\BibitemShut {NoStop}%
\bibitem [{\citenamefont {Broedersz}\ \emph {et~al.}(2011)\citenamefont
  {Broedersz}, \citenamefont {Mao}, \citenamefont {Lubensky},\ and\
  \citenamefont {MacKintosh}}]{chase_nature_physics_2011}%
  \BibitemOpen
  \bibfield  {author} {\bibinfo {author} {\bibfnamefont {C.~P.}\ \bibnamefont
  {Broedersz}}, \bibinfo {author} {\bibfnamefont {X.}~\bibnamefont {Mao}},
  \bibinfo {author} {\bibfnamefont {T.~C.}\ \bibnamefont {Lubensky}},\ and\
  \bibinfo {author} {\bibfnamefont {F.~C.}\ \bibnamefont {MacKintosh}},\
  }\bibfield  {title} {\bibinfo {title} {Criticality and isostaticity in fibre
  networks},\ }\href {https://doi.org/10.1038/nphys2127} {\bibfield  {journal}
  {\bibinfo  {journal} {Nature Physics}\ }\textbf {\bibinfo {volume} {7}},\
  \bibinfo {pages} {983} (\bibinfo {year} {2011})}\BibitemShut {NoStop}%
\bibitem [{\citenamefont {D{\"u}ring}\ \emph {et~al.}(2013)\citenamefont
  {D{\"u}ring}, \citenamefont {Lerner},\ and\ \citenamefont
  {Wyart}}]{phonon_gap_2012}%
  \BibitemOpen
  \bibfield  {author} {\bibinfo {author} {\bibfnamefont {G.}~\bibnamefont
  {D{\"u}ring}}, \bibinfo {author} {\bibfnamefont {E.}~\bibnamefont {Lerner}},\
  and\ \bibinfo {author} {\bibfnamefont {M.}~\bibnamefont {Wyart}},\ }\bibfield
   {title} {\bibinfo {title} {Phonon gap and localization lengths in floppy
  materials},\ }\href {https://doi.org/10.1039/C2SM25878A} {\bibfield
  {journal} {\bibinfo  {journal} {Soft Matter}\ }\textbf {\bibinfo {volume}
  {9}},\ \bibinfo {pages} {146} (\bibinfo {year} {2013})}\BibitemShut {NoStop}%
\bibitem [{\citenamefont {Mao}\ \emph {et~al.}(2013{\natexlab{a}})\citenamefont
  {Mao}, \citenamefont {Stenull},\ and\ \citenamefont
  {Lubensky}}]{Xiaoming_emt_2013}%
  \BibitemOpen
  \bibfield  {author} {\bibinfo {author} {\bibfnamefont {X.}~\bibnamefont
  {Mao}}, \bibinfo {author} {\bibfnamefont {O.}~\bibnamefont {Stenull}},\ and\
  \bibinfo {author} {\bibfnamefont {T.~C.}\ \bibnamefont {Lubensky}},\
  }\bibfield  {title} {\bibinfo {title} {Effective-medium theory of a
  filamentous triangular lattice},\ }\href
  {https://doi.org/10.1103/PhysRevE.87.042601} {\bibfield  {journal} {\bibinfo
  {journal} {Phys. Rev. E}\ }\textbf {\bibinfo {volume} {87}},\ \bibinfo
  {pages} {042601} (\bibinfo {year} {2013}{\natexlab{a}})}\BibitemShut
  {NoStop}%
\bibitem [{\citenamefont {Mao}\ \emph {et~al.}(2013{\natexlab{b}})\citenamefont
  {Mao}, \citenamefont {Stenull},\ and\ \citenamefont
  {Lubensky}}]{Xiaoming_kagome_2013}%
  \BibitemOpen
  \bibfield  {author} {\bibinfo {author} {\bibfnamefont {X.}~\bibnamefont
  {Mao}}, \bibinfo {author} {\bibfnamefont {O.}~\bibnamefont {Stenull}},\ and\
  \bibinfo {author} {\bibfnamefont {T.~C.}\ \bibnamefont {Lubensky}},\
  }\bibfield  {title} {\bibinfo {title} {Elasticity of a filamentous kagome
  lattice},\ }\href {https://doi.org/10.1103/PhysRevE.87.042602} {\bibfield
  {journal} {\bibinfo  {journal} {Phys. Rev. E}\ }\textbf {\bibinfo {volume}
  {87}},\ \bibinfo {pages} {042602} (\bibinfo {year}
  {2013}{\natexlab{b}})}\BibitemShut {NoStop}%
\bibitem [{\citenamefont {Lerner}(2018)}]{sss_epje_2018}%
  \BibitemOpen
  \bibfield  {author} {\bibinfo {author} {\bibfnamefont {E.}~\bibnamefont
  {Lerner}},\ }\bibfield  {title} {\bibinfo {title} {Quasilocalized states of
  self stress in packing-derived networks},\ }\href
  {https://doi.org/10.1140/epje/i2018-11705-9} {\bibfield  {journal} {\bibinfo
  {journal} {Eur. Phys. J. E}\ }\textbf {\bibinfo {volume} {41}},\ \bibinfo
  {pages} {93} (\bibinfo {year} {2018})}\BibitemShut {NoStop}%
\bibitem [{\citenamefont {Lerner}\ \emph {et~al.}(2014)\citenamefont {Lerner},
  \citenamefont {DeGiuli}, \citenamefont {During},\ and\ \citenamefont
  {Wyart}}]{breakdown}%
  \BibitemOpen
  \bibfield  {author} {\bibinfo {author} {\bibfnamefont {E.}~\bibnamefont
  {Lerner}}, \bibinfo {author} {\bibfnamefont {E.}~\bibnamefont {DeGiuli}},
  \bibinfo {author} {\bibfnamefont {G.}~\bibnamefont {During}},\ and\ \bibinfo
  {author} {\bibfnamefont {M.}~\bibnamefont {Wyart}},\ }\bibfield  {title}
  {\bibinfo {title} {Breakdown of continuum elasticity in amorphous solids},\
  }\href {https://doi.org/10.1039/C4SM00311J} {\bibfield  {journal} {\bibinfo
  {journal} {Soft Matter}\ }\textbf {\bibinfo {volume} {10}},\ \bibinfo {pages}
  {5085} (\bibinfo {year} {2014})}\BibitemShut {NoStop}%
\bibitem [{\citenamefont {Baumgarten}\ \emph {et~al.}(2017)\citenamefont
  {Baumgarten}, \citenamefont {V\aa{}gberg},\ and\ \citenamefont
  {Tighe}}]{brian_prl_2017}%
  \BibitemOpen
  \bibfield  {author} {\bibinfo {author} {\bibfnamefont {K.}~\bibnamefont
  {Baumgarten}}, \bibinfo {author} {\bibfnamefont {D.}~\bibnamefont
  {V\aa{}gberg}},\ and\ \bibinfo {author} {\bibfnamefont {B.~P.}\ \bibnamefont
  {Tighe}},\ }\bibfield  {title} {\bibinfo {title} {Nonlocal elasticity near
  jamming in frictionless soft spheres},\ }\href
  {https://doi.org/10.1103/PhysRevLett.118.098001} {\bibfield  {journal}
  {\bibinfo  {journal} {Phys. Rev. Lett.}\ }\textbf {\bibinfo {volume} {118}},\
  \bibinfo {pages} {098001} (\bibinfo {year} {2017})}\BibitemShut {NoStop}%
\bibitem [{\citenamefont {Gonz\'alez-L\'opez}\ \emph
  {et~al.}(2023)\citenamefont {Gonz\'alez-L\'opez}, \citenamefont
  {Bouchbinder},\ and\ \citenamefont {Lerner}}]{chi_paper_2023}%
  \BibitemOpen
  \bibfield  {author} {\bibinfo {author} {\bibfnamefont {K.}~\bibnamefont
  {Gonz\'alez-L\'opez}}, \bibinfo {author} {\bibfnamefont {E.}~\bibnamefont
  {Bouchbinder}},\ and\ \bibinfo {author} {\bibfnamefont {E.}~\bibnamefont
  {Lerner}},\ }\bibfield  {title} {\bibinfo {title} {Variability of mesoscopic
  mechanical disorder in disordered solids},\ }\href
  {https://doi.org/https://doi.org/10.1016/j.jnoncrysol.2023.122137} {\bibfield
   {journal} {\bibinfo  {journal} {J. Non-Cryst. Solids.}\ }\textbf {\bibinfo
  {volume} {604}},\ \bibinfo {pages} {122137} (\bibinfo {year}
  {2023})}\BibitemShut {NoStop}%
\bibitem [{\citenamefont {Giannini}\ \emph {et~al.}(2024)\citenamefont
  {Giannini}, \citenamefont {Lerner}, \citenamefont {Zamponi},\ and\
  \citenamefont {Manning}}]{giannini2024scaling}%
  \BibitemOpen
  \bibfield  {author} {\bibinfo {author} {\bibfnamefont {J.~A.}\ \bibnamefont
  {Giannini}}, \bibinfo {author} {\bibfnamefont {E.}~\bibnamefont {Lerner}},
  \bibinfo {author} {\bibfnamefont {F.}~\bibnamefont {Zamponi}},\ and\ \bibinfo
  {author} {\bibfnamefont {M.~L.}\ \bibnamefont {Manning}},\ }\bibfield
  {title} {\bibinfo {title} {{Scaling regimes and fluctuations of observables
  in computer glasses approaching the unjamming transition}},\ }\href
  {https://doi.org/10.1063/5.0176713} {\bibfield  {journal} {\bibinfo
  {journal} {J. Chem. Phys.}\ }\textbf {\bibinfo {volume} {160}},\ \bibinfo
  {pages} {034502} (\bibinfo {year} {2024})}\BibitemShut {NoStop}%
\bibitem [{\citenamefont {MacKintosh}\ \emph {et~al.}(1995)\citenamefont
  {MacKintosh}, \citenamefont {K\"as},\ and\ \citenamefont
  {Janmey}}]{PhysRevLett.75.4425}%
  \BibitemOpen
  \bibfield  {author} {\bibinfo {author} {\bibfnamefont {F.~C.}\ \bibnamefont
  {MacKintosh}}, \bibinfo {author} {\bibfnamefont {J.}~\bibnamefont {K\"as}},\
  and\ \bibinfo {author} {\bibfnamefont {P.~A.}\ \bibnamefont {Janmey}},\
  }\bibfield  {title} {\bibinfo {title} {Elasticity of semiflexible biopolymer
  networks},\ }\href {https://doi.org/10.1103/PhysRevLett.75.4425} {\bibfield
  {journal} {\bibinfo  {journal} {Phys. Rev. Lett.}\ }\textbf {\bibinfo
  {volume} {75}},\ \bibinfo {pages} {4425} (\bibinfo {year}
  {1995})}\BibitemShut {NoStop}%
\bibitem [{\citenamefont {Lindstr\"om}\ \emph {et~al.}(2010)\citenamefont
  {Lindstr\"om}, \citenamefont {Vader}, \citenamefont {Kulachenko},\ and\
  \citenamefont {Weitz}}]{PhysRevE.82.051905}%
  \BibitemOpen
  \bibfield  {author} {\bibinfo {author} {\bibfnamefont {S.~B.}\ \bibnamefont
  {Lindstr\"om}}, \bibinfo {author} {\bibfnamefont {D.~A.}\ \bibnamefont
  {Vader}}, \bibinfo {author} {\bibfnamefont {A.}~\bibnamefont {Kulachenko}},\
  and\ \bibinfo {author} {\bibfnamefont {D.~A.}\ \bibnamefont {Weitz}},\
  }\bibfield  {title} {\bibinfo {title} {Biopolymer network geometries:
  Characterization, regeneration, and elastic properties},\ }\href
  {https://doi.org/10.1103/PhysRevE.82.051905} {\bibfield  {journal} {\bibinfo
  {journal} {Phys. Rev. E}\ }\textbf {\bibinfo {volume} {82}},\ \bibinfo
  {pages} {051905} (\bibinfo {year} {2010})}\BibitemShut {NoStop}%
\bibitem [{\citenamefont {Head}\ \emph {et~al.}(2003)\citenamefont {Head},
  \citenamefont {Levine},\ and\ \citenamefont {MacKintosh}}]{fred_prl_2003}%
  \BibitemOpen
  \bibfield  {author} {\bibinfo {author} {\bibfnamefont {D.~A.}\ \bibnamefont
  {Head}}, \bibinfo {author} {\bibfnamefont {A.~J.}\ \bibnamefont {Levine}},\
  and\ \bibinfo {author} {\bibfnamefont {F.~C.}\ \bibnamefont {MacKintosh}},\
  }\bibfield  {title} {\bibinfo {title} {Deformation of cross-linked
  semiflexible polymer networks},\ }\href
  {https://doi.org/10.1103/PhysRevLett.91.108102} {\bibfield  {journal}
  {\bibinfo  {journal} {Phys. Rev. Lett.}\ }\textbf {\bibinfo {volume} {91}},\
  \bibinfo {pages} {108102} (\bibinfo {year} {2003})}\BibitemShut {NoStop}%
\bibitem [{\citenamefont {Broedersz}\ and\ \citenamefont
  {MacKintosh}(2014)}]{RevModPhys.86.995}%
  \BibitemOpen
  \bibfield  {author} {\bibinfo {author} {\bibfnamefont {C.~P.}\ \bibnamefont
  {Broedersz}}\ and\ \bibinfo {author} {\bibfnamefont {F.~C.}\ \bibnamefont
  {MacKintosh}},\ }\bibfield  {title} {\bibinfo {title} {Modeling semiflexible
  polymer networks},\ }\href {https://doi.org/10.1103/RevModPhys.86.995}
  {\bibfield  {journal} {\bibinfo  {journal} {Rev. Mod. Phys.}\ }\textbf
  {\bibinfo {volume} {86}},\ \bibinfo {pages} {995} (\bibinfo {year}
  {2014})}\BibitemShut {NoStop}%
\bibitem [{\citenamefont {Shivers}\ \emph {et~al.}(2019)\citenamefont
  {Shivers}, \citenamefont {Arzash}, \citenamefont {Sharma},\ and\
  \citenamefont {MacKintosh}}]{mackintosh_prl_2019}%
  \BibitemOpen
  \bibfield  {author} {\bibinfo {author} {\bibfnamefont {J.~L.}\ \bibnamefont
  {Shivers}}, \bibinfo {author} {\bibfnamefont {S.}~\bibnamefont {Arzash}},
  \bibinfo {author} {\bibfnamefont {A.}~\bibnamefont {Sharma}},\ and\ \bibinfo
  {author} {\bibfnamefont {F.~C.}\ \bibnamefont {MacKintosh}},\ }\bibfield
  {title} {\bibinfo {title} {Scaling theory for mechanical critical behavior in
  fiber networks},\ }\href {https://doi.org/10.1103/PhysRevLett.122.188003}
  {\bibfield  {journal} {\bibinfo  {journal} {Phys. Rev. Lett.}\ }\textbf
  {\bibinfo {volume} {122}},\ \bibinfo {pages} {188003} (\bibinfo {year}
  {2019})}\BibitemShut {NoStop}%
\bibitem [{\citenamefont {D\"uring}\ \emph {et~al.}(2014)\citenamefont
  {D\"uring}, \citenamefont {Lerner},\ and\ \citenamefont
  {Wyart}}]{gustavo_pre_2014}%
  \BibitemOpen
  \bibfield  {author} {\bibinfo {author} {\bibfnamefont {G.}~\bibnamefont
  {D\"uring}}, \bibinfo {author} {\bibfnamefont {E.}~\bibnamefont {Lerner}},\
  and\ \bibinfo {author} {\bibfnamefont {M.}~\bibnamefont {Wyart}},\ }\bibfield
   {title} {\bibinfo {title} {Length scales and self-organization in dense
  suspension flows},\ }\href {https://doi.org/10.1103/PhysRevE.89.022305}
  {\bibfield  {journal} {\bibinfo  {journal} {Phys. Rev. E}\ }\textbf {\bibinfo
  {volume} {89}},\ \bibinfo {pages} {022305} (\bibinfo {year}
  {2014})}\BibitemShut {NoStop}%
\bibitem [{\citenamefont {Feng}\ \emph {et~al.}(2016)\citenamefont {Feng},
  \citenamefont {Levine}, \citenamefont {Mao},\ and\ \citenamefont
  {Sander}}]{feng_soft_matter_2016}%
  \BibitemOpen
  \bibfield  {author} {\bibinfo {author} {\bibfnamefont {J.}~\bibnamefont
  {Feng}}, \bibinfo {author} {\bibfnamefont {H.}~\bibnamefont {Levine}},
  \bibinfo {author} {\bibfnamefont {X.}~\bibnamefont {Mao}},\ and\ \bibinfo
  {author} {\bibfnamefont {L.~M.}\ \bibnamefont {Sander}},\ }\bibfield  {title}
  {\bibinfo {title} {Nonlinear elasticity of disordered fiber networks},\
  }\href {https://doi.org/10.1039/C5SM01856K} {\bibfield  {journal} {\bibinfo
  {journal} {Soft Matter}\ }\textbf {\bibinfo {volume} {12}},\ \bibinfo {pages}
  {1419} (\bibinfo {year} {2016})}\BibitemShut {NoStop}%
\bibitem [{\citenamefont {Sharma}\ \emph {et~al.}(2016)\citenamefont {Sharma},
  \citenamefont {Licup}, \citenamefont {Jansen}, \citenamefont {Rens},
  \citenamefont {Sheinman}, \citenamefont {Koenderink},\ and\ \citenamefont
  {MacKintosh}}]{robbie_nature_physics_2016}%
  \BibitemOpen
  \bibfield  {author} {\bibinfo {author} {\bibfnamefont {A.}~\bibnamefont
  {Sharma}}, \bibinfo {author} {\bibfnamefont {A.~J.}\ \bibnamefont {Licup}},
  \bibinfo {author} {\bibfnamefont {K.~A.}\ \bibnamefont {Jansen}}, \bibinfo
  {author} {\bibfnamefont {R.}~\bibnamefont {Rens}}, \bibinfo {author}
  {\bibfnamefont {M.}~\bibnamefont {Sheinman}}, \bibinfo {author}
  {\bibfnamefont {G.~H.}\ \bibnamefont {Koenderink}},\ and\ \bibinfo {author}
  {\bibfnamefont {F.~C.}\ \bibnamefont {MacKintosh}},\ }\bibfield  {title}
  {\bibinfo {title} {Strain-controlled criticality governs the nonlinear
  mechanics of fibre networks},\ }\href {https://doi.org/10.1038/nphys3628}
  {\bibfield  {journal} {\bibinfo  {journal} {Nature Physics}\ }\textbf
  {\bibinfo {volume} {12}},\ \bibinfo {pages} {584} (\bibinfo {year}
  {2016})}\BibitemShut {NoStop}%
\bibitem [{\citenamefont {Licup}\ \emph {et~al.}(2016)\citenamefont {Licup},
  \citenamefont {Sharma},\ and\ \citenamefont
  {MacKintosh}}]{mackintosh_pre_2016}%
  \BibitemOpen
  \bibfield  {author} {\bibinfo {author} {\bibfnamefont {A.~J.}\ \bibnamefont
  {Licup}}, \bibinfo {author} {\bibfnamefont {A.}~\bibnamefont {Sharma}},\ and\
  \bibinfo {author} {\bibfnamefont {F.~C.}\ \bibnamefont {MacKintosh}},\
  }\bibfield  {title} {\bibinfo {title} {Elastic regimes of subisostatic
  athermal fiber networks},\ }\href
  {https://doi.org/10.1103/PhysRevE.93.012407} {\bibfield  {journal} {\bibinfo
  {journal} {Phys. Rev. E}\ }\textbf {\bibinfo {volume} {93}},\ \bibinfo
  {pages} {012407} (\bibinfo {year} {2016})}\BibitemShut {NoStop}%
\bibitem [{\citenamefont {Rens}\ \emph {et~al.}(2018)\citenamefont {Rens},
  \citenamefont {Villarroel}, \citenamefont {D\"uring},\ and\ \citenamefont
  {Lerner}}]{robbie_pre_2018}%
  \BibitemOpen
  \bibfield  {author} {\bibinfo {author} {\bibfnamefont {R.}~\bibnamefont
  {Rens}}, \bibinfo {author} {\bibfnamefont {C.}~\bibnamefont {Villarroel}},
  \bibinfo {author} {\bibfnamefont {G.}~\bibnamefont {D\"uring}},\ and\
  \bibinfo {author} {\bibfnamefont {E.}~\bibnamefont {Lerner}},\ }\bibfield
  {title} {\bibinfo {title} {Micromechanical theory of strain stiffening of
  biopolymer networks},\ }\href {https://doi.org/10.1103/PhysRevE.98.062411}
  {\bibfield  {journal} {\bibinfo  {journal} {Phys. Rev. E}\ }\textbf {\bibinfo
  {volume} {98}},\ \bibinfo {pages} {062411} (\bibinfo {year}
  {2018})}\BibitemShut {NoStop}%
\bibitem [{\citenamefont {Merkel}\ \emph {et~al.}(2019)\citenamefont {Merkel},
  \citenamefont {Baumgarten}, \citenamefont {Tighe},\ and\ \citenamefont
  {Manning}}]{merkel_pnas_2019}%
  \BibitemOpen
  \bibfield  {author} {\bibinfo {author} {\bibfnamefont {M.}~\bibnamefont
  {Merkel}}, \bibinfo {author} {\bibfnamefont {K.}~\bibnamefont {Baumgarten}},
  \bibinfo {author} {\bibfnamefont {B.~P.}\ \bibnamefont {Tighe}},\ and\
  \bibinfo {author} {\bibfnamefont {M.~L.}\ \bibnamefont {Manning}},\
  }\bibfield  {title} {\bibinfo {title} {A minimal-length approach unifies
  rigidity in underconstrained materials},\ }\href
  {https://doi.org/10.1073/pnas.1815436116} {\bibfield  {journal} {\bibinfo
  {journal} {Proc. Natl. Acad. Sci. U.S.A.}\ }\textbf {\bibinfo {volume}
  {116}},\ \bibinfo {pages} {6560} (\bibinfo {year} {2019})}\BibitemShut
  {NoStop}%
\bibitem [{\citenamefont {Shivers}\ \emph {et~al.}(2023)\citenamefont
  {Shivers}, \citenamefont {Sharma},\ and\ \citenamefont
  {MacKintosh}}]{fred_prl_2023}%
  \BibitemOpen
  \bibfield  {author} {\bibinfo {author} {\bibfnamefont {J.~L.}\ \bibnamefont
  {Shivers}}, \bibinfo {author} {\bibfnamefont {A.}~\bibnamefont {Sharma}},\
  and\ \bibinfo {author} {\bibfnamefont {F.~C.}\ \bibnamefont {MacKintosh}},\
  }\bibfield  {title} {\bibinfo {title} {Strain-controlled critical slowing
  down in the rheology of disordered networks},\ }\href
  {https://doi.org/10.1103/PhysRevLett.131.178201} {\bibfield  {journal}
  {\bibinfo  {journal} {Phys. Rev. Lett.}\ }\textbf {\bibinfo {volume} {131}},\
  \bibinfo {pages} {178201} (\bibinfo {year} {2023})}\BibitemShut {NoStop}%
\bibitem [{\citenamefont {Lerner}\ and\ \citenamefont
  {Bouchbinder}(2023{\natexlab{a}})}]{strain_stiffening_2023}%
  \BibitemOpen
  \bibfield  {author} {\bibinfo {author} {\bibfnamefont {E.}~\bibnamefont
  {Lerner}}\ and\ \bibinfo {author} {\bibfnamefont {E.}~\bibnamefont
  {Bouchbinder}},\ }\bibfield  {title} {\bibinfo {title} {Scaling theory of
  critical strain-stiffening in disordered elastic networks},\ }\href
  {https://doi.org/https://doi.org/10.1016/j.eml.2023.102104} {\bibfield
  {journal} {\bibinfo  {journal} {Extreme Mechanics Letters}\ }\textbf
  {\bibinfo {volume} {65}},\ \bibinfo {pages} {102104} (\bibinfo {year}
  {2023}{\natexlab{a}})}\BibitemShut {NoStop}%
\bibitem [{\citenamefont {Rens}\ and\ \citenamefont
  {Lerner}(2019)}]{robbie_epje_2019}%
  \BibitemOpen
  \bibfield  {author} {\bibinfo {author} {\bibfnamefont {R.}~\bibnamefont
  {Rens}}\ and\ \bibinfo {author} {\bibfnamefont {E.}~\bibnamefont {Lerner}},\
  }\bibfield  {title} {\bibinfo {title} {Rigidity and auxeticity transitions in
  networks with strong bond-bending interactions},\ }\href
  {https://doi.org/10.1140/epje/i2019-11888-5} {\bibfield  {journal} {\bibinfo
  {journal} {Eur. Phys. J. E}\ }\textbf {\bibinfo {volume} {42}},\ \bibinfo
  {pages} {114} (\bibinfo {year} {2019})}\BibitemShut {NoStop}%
\bibitem [{\citenamefont {Kapteijns}\ \emph {et~al.}(2021)\citenamefont
  {Kapteijns}, \citenamefont {Bouchbinder},\ and\ \citenamefont
  {Lerner}}]{phonon_width_2}%
  \BibitemOpen
  \bibfield  {author} {\bibinfo {author} {\bibfnamefont {G.}~\bibnamefont
  {Kapteijns}}, \bibinfo {author} {\bibfnamefont {E.}~\bibnamefont
  {Bouchbinder}},\ and\ \bibinfo {author} {\bibfnamefont {E.}~\bibnamefont
  {Lerner}},\ }\bibfield  {title} {\bibinfo {title} {Unified quantifier of
  mechanical disorder in solids},\ }\href
  {https://doi.org/10.1103/PhysRevE.104.035001} {\bibfield  {journal} {\bibinfo
   {journal} {Phys. Rev. E}\ }\textbf {\bibinfo {volume} {104}},\ \bibinfo
  {pages} {035001} (\bibinfo {year} {2021})}\BibitemShut {NoStop}%
\bibitem [{\citenamefont {Jacobs}\ and\ \citenamefont
  {Thorpe}(1996)}]{Thorpe_pre_1996}%
  \BibitemOpen
  \bibfield  {author} {\bibinfo {author} {\bibfnamefont {D.~J.}\ \bibnamefont
  {Jacobs}}\ and\ \bibinfo {author} {\bibfnamefont {M.~F.}\ \bibnamefont
  {Thorpe}},\ }\bibfield  {title} {\bibinfo {title} {Generic rigidity
  percolation in two dimensions},\ }\href
  {https://doi.org/10.1103/PhysRevE.53.3682} {\bibfield  {journal} {\bibinfo
  {journal} {Phys. Rev. E}\ }\textbf {\bibinfo {volume} {53}},\ \bibinfo
  {pages} {3682} (\bibinfo {year} {1996})}\BibitemShut {NoStop}%
\bibitem [{\citenamefont {Ellenbroek}\ \emph {et~al.}(2015)\citenamefont
  {Ellenbroek}, \citenamefont {Hagh}, \citenamefont {Kumar}, \citenamefont
  {Thorpe},\ and\ \citenamefont {van Hecke}}]{ellenbroek_rigidity_prl_2015}%
  \BibitemOpen
  \bibfield  {author} {\bibinfo {author} {\bibfnamefont {W.~G.}\ \bibnamefont
  {Ellenbroek}}, \bibinfo {author} {\bibfnamefont {V.~F.}\ \bibnamefont
  {Hagh}}, \bibinfo {author} {\bibfnamefont {A.}~\bibnamefont {Kumar}},
  \bibinfo {author} {\bibfnamefont {M.~F.}\ \bibnamefont {Thorpe}},\ and\
  \bibinfo {author} {\bibfnamefont {M.}~\bibnamefont {van Hecke}},\ }\bibfield
  {title} {\bibinfo {title} {Rigidity loss in disordered systems: Three
  scenarios},\ }\href {https://doi.org/10.1103/PhysRevLett.114.135501}
  {\bibfield  {journal} {\bibinfo  {journal} {Phys. Rev. Lett.}\ }\textbf
  {\bibinfo {volume} {114}},\ \bibinfo {pages} {135501} (\bibinfo {year}
  {2015})}\BibitemShut {NoStop}%
\bibitem [{\citenamefont {Lerner}\ and\ \citenamefont
  {Bouchbinder}(2023{\natexlab{b}})}]{anomalous_elasticity_soft_matter_2023}%
  \BibitemOpen
  \bibfield  {author} {\bibinfo {author} {\bibfnamefont {E.}~\bibnamefont
  {Lerner}}\ and\ \bibinfo {author} {\bibfnamefont {E.}~\bibnamefont
  {Bouchbinder}},\ }\bibfield  {title} {\bibinfo {title} {Anomalous linear
  elasticity of disordered networks},\ }\href
  {https://doi.org/10.1039/D2SM01253G} {\bibfield  {journal} {\bibinfo
  {journal} {Soft Matter}\ }\textbf {\bibinfo {volume} {19}},\ \bibinfo {pages}
  {1076} (\bibinfo {year} {2023}{\natexlab{b}})}\BibitemShut {NoStop}%
\bibitem [{\citenamefont {Maloney}\ and\ \citenamefont
  {Lema\^{\i}tre}(2004)}]{lemaitre2004_avalanches}%
  \BibitemOpen
  \bibfield  {author} {\bibinfo {author} {\bibfnamefont {C.}~\bibnamefont
  {Maloney}}\ and\ \bibinfo {author} {\bibfnamefont {A.}~\bibnamefont
  {Lema\^{\i}tre}},\ }\bibfield  {title} {\bibinfo {title} {Subextensive
  scaling in the athermal, quasistatic limit of amorphous matter in plastic
  shear flow},\ }\href {https://doi.org/10.1103/PhysRevLett.93.016001}
  {\bibfield  {journal} {\bibinfo  {journal} {Phys. Rev. Lett.}\ }\textbf
  {\bibinfo {volume} {93}},\ \bibinfo {pages} {016001} (\bibinfo {year}
  {2004})}\BibitemShut {NoStop}%
\bibitem [{\citenamefont {Bitzek}\ \emph {et~al.}(2006)\citenamefont {Bitzek},
  \citenamefont {Koskinen}, \citenamefont {G\"ahler}, \citenamefont {Moseler},\
  and\ \citenamefont {Gumbsch}}]{fire}%
  \BibitemOpen
  \bibfield  {author} {\bibinfo {author} {\bibfnamefont {E.}~\bibnamefont
  {Bitzek}}, \bibinfo {author} {\bibfnamefont {P.}~\bibnamefont {Koskinen}},
  \bibinfo {author} {\bibfnamefont {F.}~\bibnamefont {G\"ahler}}, \bibinfo
  {author} {\bibfnamefont {M.}~\bibnamefont {Moseler}},\ and\ \bibinfo {author}
  {\bibfnamefont {P.}~\bibnamefont {Gumbsch}},\ }\bibfield  {title} {\bibinfo
  {title} {Structural relaxation made simple},\ }\href
  {https://doi.org/10.1103/PhysRevLett.97.170201} {\bibfield  {journal}
  {\bibinfo  {journal} {Phys. Rev. Lett.}\ }\textbf {\bibinfo {volume} {97}},\
  \bibinfo {pages} {170201} (\bibinfo {year} {2006})}\BibitemShut {NoStop}%
\bibitem [{\citenamefont {Rens}\ \emph {et~al.}(2016)\citenamefont {Rens},
  \citenamefont {Vahabi}, \citenamefont {Licup}, \citenamefont {MacKintosh},\
  and\ \citenamefont {Sharma}}]{Rens_JPCB_2016}%
  \BibitemOpen
  \bibfield  {author} {\bibinfo {author} {\bibfnamefont {R.}~\bibnamefont
  {Rens}}, \bibinfo {author} {\bibfnamefont {M.}~\bibnamefont {Vahabi}},
  \bibinfo {author} {\bibfnamefont {A.~J.}\ \bibnamefont {Licup}}, \bibinfo
  {author} {\bibfnamefont {F.~C.}\ \bibnamefont {MacKintosh}},\ and\ \bibinfo
  {author} {\bibfnamefont {A.}~\bibnamefont {Sharma}},\ }\bibfield  {title}
  {\bibinfo {title} {Nonlinear mechanics of athermal branched biopolymer
  networks},\ }\href {https://doi.org/10.1021/acs.jpcb.6b00259} {\bibfield
  {journal} {\bibinfo  {journal} {J. Phys. Chem. B}\ }\textbf {\bibinfo
  {volume} {120}},\ \bibinfo {pages} {5831} (\bibinfo {year}
  {2016})}\BibitemShut {NoStop}%
\bibitem [{\citenamefont {Wyart}\ \emph {et~al.}(2005)\citenamefont {Wyart},
  \citenamefont {Silbert}, \citenamefont {Nagel},\ and\ \citenamefont
  {Witten}}]{matthieu_PRE_2005}%
  \BibitemOpen
  \bibfield  {author} {\bibinfo {author} {\bibfnamefont {M.}~\bibnamefont
  {Wyart}}, \bibinfo {author} {\bibfnamefont {L.~E.}\ \bibnamefont {Silbert}},
  \bibinfo {author} {\bibfnamefont {S.~R.}\ \bibnamefont {Nagel}},\ and\
  \bibinfo {author} {\bibfnamefont {T.~A.}\ \bibnamefont {Witten}},\ }\bibfield
   {title} {\bibinfo {title} {Effects of compression on the vibrational modes
  of marginally jammed solids},\ }\href
  {https://doi.org/10.1103/PhysRevE.72.051306} {\bibfield  {journal} {\bibinfo
  {journal} {Phys. Rev. E}\ }\textbf {\bibinfo {volume} {72}},\ \bibinfo
  {pages} {051306} (\bibinfo {year} {2005})}\BibitemShut {NoStop}%
\bibitem [{\citenamefont {Yan}\ \emph {et~al.}(2016)\citenamefont {Yan},
  \citenamefont {DeGiuli},\ and\ \citenamefont
  {Wyart}}]{new_variational_argument_epl_2016}%
  \BibitemOpen
  \bibfield  {author} {\bibinfo {author} {\bibfnamefont {L.}~\bibnamefont
  {Yan}}, \bibinfo {author} {\bibfnamefont {E.}~\bibnamefont {DeGiuli}},\ and\
  \bibinfo {author} {\bibfnamefont {M.}~\bibnamefont {Wyart}},\ }\bibfield
  {title} {\bibinfo {title} {On variational arguments for vibrational modes
  near jamming},\ }\href {http://stacks.iop.org/0295-5075/114/i=2/a=26003}
  {\bibfield  {journal} {\bibinfo  {journal} {Europhys. Lett.}\ }\textbf
  {\bibinfo {volume} {114}},\ \bibinfo {pages} {26003} (\bibinfo {year}
  {2016})}\BibitemShut {NoStop}%
\bibitem [{\citenamefont {Ellenbroek}\ \emph {et~al.}(2006)\citenamefont
  {Ellenbroek}, \citenamefont {Somfai}, \citenamefont {van Hecke},\ and\
  \citenamefont {van Saarloos}}]{wouter_prl_2006}%
  \BibitemOpen
  \bibfield  {author} {\bibinfo {author} {\bibfnamefont {W.~G.}\ \bibnamefont
  {Ellenbroek}}, \bibinfo {author} {\bibfnamefont {E.}~\bibnamefont {Somfai}},
  \bibinfo {author} {\bibfnamefont {M.}~\bibnamefont {van Hecke}},\ and\
  \bibinfo {author} {\bibfnamefont {W.}~\bibnamefont {van Saarloos}},\
  }\bibfield  {title} {\bibinfo {title} {Critical scaling in linear response of
  frictionless granular packings near jamming},\ }\href
  {https://doi.org/10.1103/PhysRevLett.97.258001} {\bibfield  {journal}
  {\bibinfo  {journal} {Phys. Rev. Lett.}\ }\textbf {\bibinfo {volume} {97}},\
  \bibinfo {pages} {258001} (\bibinfo {year} {2006})}\BibitemShut {NoStop}%
\bibitem [{\citenamefont {Silbert}\ \emph {et~al.}(2005)\citenamefont
  {Silbert}, \citenamefont {Liu},\ and\ \citenamefont
  {Nagel}}]{Silbert_prl_2005}%
  \BibitemOpen
  \bibfield  {author} {\bibinfo {author} {\bibfnamefont {L.~E.}\ \bibnamefont
  {Silbert}}, \bibinfo {author} {\bibfnamefont {A.~J.}\ \bibnamefont {Liu}},\
  and\ \bibinfo {author} {\bibfnamefont {S.~R.}\ \bibnamefont {Nagel}},\
  }\bibfield  {title} {\bibinfo {title} {Vibrations and diverging length scales
  near the unjamming transition},\ }\href
  {https://doi.org/10.1103/PhysRevLett.95.098301} {\bibfield  {journal}
  {\bibinfo  {journal} {Phys. Rev. Lett.}\ }\textbf {\bibinfo {volume} {95}},\
  \bibinfo {pages} {098301} (\bibinfo {year} {2005})}\BibitemShut {NoStop}%
\bibitem [{\citenamefont {Rens}(2019)}]{robbie_thesis}%
  \BibitemOpen
  \bibfield  {author} {\bibinfo {author} {\bibfnamefont {R.}~\bibnamefont
  {Rens}},\ }\href
  {https://dare.uva.nl/search?identifier=6ae60739-28c6-4318-aab9-939a0dacc4bc}
  {\emph {\bibinfo {title} {Theory of rigidity transitions in disordered
  materials}}}\ (\bibinfo  {publisher} {PhD thesis, Univeristy of Amsterdam,
  the Netherlands},\ \bibinfo {year} {2019})\BibitemShut {NoStop}%
\bibitem [{\citenamefont {Vermeulen}\ \emph {et~al.}(2017)\citenamefont
  {Vermeulen}, \citenamefont {Bose}, \citenamefont {Storm},\ and\ \citenamefont
  {Ellenbroek}}]{wouter_pre_2017}%
  \BibitemOpen
  \bibfield  {author} {\bibinfo {author} {\bibfnamefont {M.~F.~J.}\
  \bibnamefont {Vermeulen}}, \bibinfo {author} {\bibfnamefont {A.}~\bibnamefont
  {Bose}}, \bibinfo {author} {\bibfnamefont {C.}~\bibnamefont {Storm}},\ and\
  \bibinfo {author} {\bibfnamefont {W.~G.}\ \bibnamefont {Ellenbroek}},\
  }\bibfield  {title} {\bibinfo {title} {Geometry and the onset of rigidity in
  a disordered network},\ }\href {https://doi.org/10.1103/PhysRevE.96.053003}
  {\bibfield  {journal} {\bibinfo  {journal} {Phys. Rev. E}\ }\textbf {\bibinfo
  {volume} {96}},\ \bibinfo {pages} {053003} (\bibinfo {year}
  {2017})}\BibitemShut {NoStop}%
\end{thebibliography}
%

\end{document}